\documentclass{IEEEoj}
\usepackage[sort, compress]{cite}
\usepackage{amsmath,amssymb,amsfonts}
\usepackage{algorithmic}
\usepackage{graphicx,color}
\usepackage{textcomp}

\usepackage{hyperref}
\usepackage{float}
 \usepackage{dblfloatfix}

\renewcommand*{\thesection}{\Roman{section}}

\usepackage{placeins}
\usepackage{times}
\usepackage{epsfig}
\usepackage{amsthm}
\usepackage{booktabs}

\usepackage{tabularx}
\usepackage{subcaption}
\usepackage{dblfloatfix}

\newtheorem{theorem}{Theorem}
\newtheorem{definition}{Definition}

\newtheorem{remark}{Remark}

\usepackage{bm}

\long\def\comment#1{}

\newfont{\bbb}{msbm10 scaled 700}

\newcommand{\Am}{{\bf A}}

\newcommand{\Dm}{{\bf D}}

\newcommand{\Fm}{{\bf F}}

\newcommand{\Id}{{\bf I}}

\newcommand{\Qm}{{\bf Q}}

\newcommand{\Wm}{{\bf W}}

\newcommand{\Xm}{{\bf X}}

\def\BibTeX{{\rm B\kern-.05em{\sc i\kern-.025em b}\kern-.08em
    T\kern-.1667em\lower.7ex\hbox{E}\kern-.125emX}}
\AtBeginDocument{\definecolor{ojcolor}{cmyk}{0.93,0.59,0.15,0.02}}
 \def\OJlogo{\vspace{-14pt}\includegraphics[height=28pt]{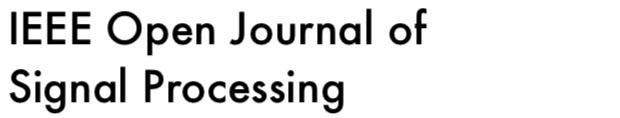}}
\begin{document}
\receiveddate{}
\reviseddate{}
\accepteddate{}
\publisheddate{}
\currentdate{}
\doiinfo{}

\title{Towards a geometric understanding of Spatio Temporal Graph Convolution Networks}

\author{Pratyusha Das, IEEE Student Member, Sarath Shekkizhar, IEEE Student Member, Antonio Ortega, IEEE Fellow}
\affil{University of Southern California, Los Angeles, CA 90089, USA}

\corresp{CORRESPONDING AUTHOR: Pratyusha Das (e-mail: daspraty@usc.edu).}
% \authornote{This work was supported by the Natural Sciences and Engineering Research Council (NSERC) of Canada.}
\markboth{Preparation of Papers for IEEE OPEN JOURNALS}{Das \textit{et al.}}

\begin{abstract}
Spatiotemporal graph convolutional networks (STGCNs) have emerged as a desirable model for \emph{skeleton}-based human action recognition. 
Despite achieving state-of-the-art performance, there is a limited understanding of the 
representations learned by these models, which hinders their application in critical and real-world settings. While layerwise analysis of CNN models has been studied in the literature, to the best of our knowledge, there exists \emph{no study} on the layerwise explainability of the embeddings learned on spatiotemporal data using STGCNs. 
In this paper, we first propose to use a local 
Dataset Graph (DS-Graph) obtained from the feature representation of input data at each layer to develop an understanding of the layer-wise embedding geometry of the STGCN. 
To do so, we develop a 
window-based dynamic time warping (DTW) method to compute the distance between data sequences with varying temporal lengths. 
To validate our findings, we have developed a layer-specific Spatiotemporal Graph Gradient-weighted Class Activation Mapping (L-STG-GradCAM) technique tailored for spatiotemporal data. 
This approach enables us to visually analyze and interpret each layer within the STGCN network.
We characterize the functions learned by each layer of the STGCN using the label smoothness 
%\cite{zhu2003semi} 
of the representation and visualize them using our L-STG-GradCAM approach. Our proposed method is generic and can yield valuable insights for STGCN architectures in different applications. However, this paper focuses on the human activity recognition task as a representative application.
Our experiments show that STGCN models learn representations that capture general human motion in their initial layers while discriminating different actions only in later layers. 
This justifies experimental observations showing that fine-tuning deeper layers works well for transfer between related tasks. 
We provide experimental evidence for different human activity datasets and advanced spatiotemporal graph networks to validate that the proposed method is general enough to analyze any STGCN model and can be useful for drawing insight into networks in various scenarios.   
We also show that noise at the input has a limited effect on label smoothness, which can help justify the robustness of STGCNs to noise.

\end{abstract}



\maketitle

\section{INTRODUCTION}
\IEEEPARstart{D}{eep} 
 learning models have led to significant advances 
in application domains, such as images and video  \cite{krizhevsky2012imagenet, he2016deep}, where data is available on a regular grid, e.g., formed by pixels. More recently, graph neural networks (GNNs) \cite{wu2020comprehensive}, and graph convolutional networks (GCNs) \cite{wu2019simplifying} have been proposed to handle \emph{data with irregular structures}, such as social networks \cite{pham2015general}, skeleton-based motion capture data (MoCap) \cite{kao2019graph}.
In this paper, we focus on spatiotemporal graph convolutional networks (STGCNs). STGCNs can efficiently handle the temporal aspects of graph data and have wide-ranging applications, including in tasks such as traffic forecasting \cite{yu2017spatio} and the recognition of actions based on skeletal data \cite{ren2020survey}.

\begin{figure*}[btp]
\begin{center}
\frame{\includegraphics[width=1\linewidth]{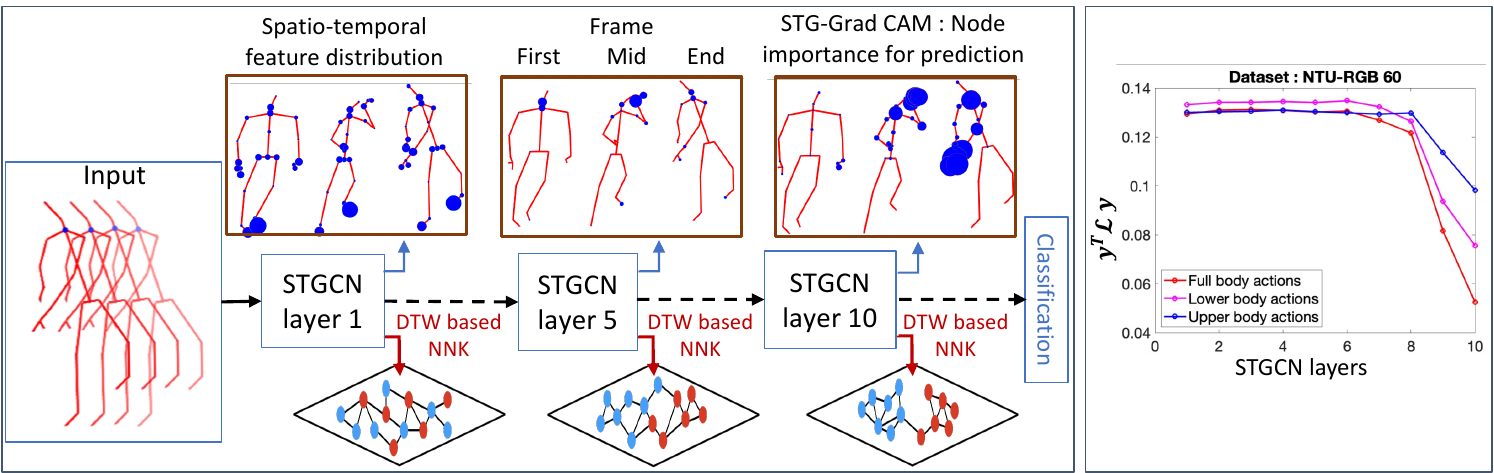}}
% \fbox{\includegraphics[width=1\linewidth]{images/figure1.pdf}\rule{0pt}{2in} \rule{.9\linewidth}{0pt}}
    
\end{center}
\caption{Proposed data-driven approach to understanding the geometry of the embedding manifold in STGCNs using windowed dynamic time warping (DTW) and Non-Negative Kernel (NNK) graphs. 
   %We construct \textit{Dataset  NNK Graphs}  where each node corresponds to an action sequence, and the weights of edges connecting two nodes are derived from pairwise distances between the features representing the corresponding sequences. Our graph-based approach allows us to compare different representations (skeleton graphs) of the same dataset, even if the representations are heterogeneous, e.g., if they correspond to network layers with different dimensions. 
   \textbf{Left:} We construct \textit{Dataset  NNK Graphs (DS-Graph)}  where each node corresponds to an action sequence, and the weights of edges connecting two nodes are derived from pairwise distances between the features representing the corresponding action sequences.  
   In this example, we show how the two classes 
   (corresponding to red and blue nodes on the DS-Graph) become more clearly separated in deeper layers of the network. We also observe the \textit{skeleton graph (S-Graph) } node importance for each action using a layerwise STG-GradCAM (the three-time slice example corresponds to a \emph{Throw} action). \textbf{Right:} For a set of spatiotemporal input action sequences, we observe the label smoothness on the \textit{DS-Graph} constructed using the features obtained for the sequences after each STGCN layer. The observed label smoothness at each layer of the STGCN network averaged over three super-classes corresponding to actions involving the upper body, lower body, and full body. In this plot, lower variation corresponds to greater smoothness. We note that the label smoothness increases in the deeper layers, in which the different actions can be classified (see DS-Graphs at the bottom of the left plot).} 
   %Also, note that lower-body actions are learned better than upper-body actions as they achieved lower values of  the Laplacian quadratic form.}
\label{fig:summary}
\end{figure*}

% Skeleton data has been widely used in human action recognition due to its view-invariant representation of pose structure, robustness against sensor noise \cite{zhang2012microsoft, cao2017realtime}, and efficiency in computation and storage \cite{li2018co,yan2017skeleton}. 
% Early approaches made use of the structure of the skeleton graph to capture the correlation between different skeleton joints for action recognition  
% \cite{kao2019graph, duvenaud2015convolutional}.
% Recently, STGCN approaches have quickly gained popularity with superior performance in human activity understanding \cite{li2019actional, liu2020disentangling, das2021symmetric}. 

When it comes to training STGCN models, there are several crucial design choices to consider, such as the architecture, optimization routine, loss function, and dataset. 
Usually, these choices interplay in intricate ways to shape the characteristics of the final model. Therefore, selecting a particular model is often primarily based on its \emph{performance} on specific datasets. While this practical perspective has led to significant advances, achieving a deeper understanding of the system is essential for ensuring safe and robust real-world deployment.

Two major approaches have been used in the literature to understand deep learning systems. 
\textit{Function approximation} methods are based on the inductive bias of the loss function \cite{ghorbani2019investigation}, the ability of the optimization to achieve good minima \cite{arora2019fine}, or consider the study of classifier margins \cite{gunasekar2018implicit}. 
\textit{Data-driven} analysis methods consider the relative position of sample data points in the representation domain %model's classification performance on different data points 
 for characterization \cite{baldock2021deep, shekkizhar2021model, SSL_geometry}. 
Data-driven approaches can provide a unified framework for understanding models because they can abstract the specific functional components. Specifically, in a data-driven approach, functions do not need to be explicitly modeled; they can be characterized implicitly using the outputs they produce. 

In this paper, we develop a \textit{data-driven} approach to achieve a better understanding of STGCN models. 
Our approach is based on a layer-wise analysis, interpretation, and visualization of the embeddings produced by the STGCN. 
Our proposed method starts by defining a \textbf{Dataset graph (DS-Graph)}, which captures the pairwise similarities between sequences in the set, represented by their embedding.  This allows us to compare models obtained with very different architectures by simply comparing the DS-Graphs they produce in their respective embedded spaces. 
While our method is widely applicable, our experiments focus on a human activity recognition task using skeleton-based data as an illustrative task to evaluate our STGCN analysis methods. 
This type of data has been widely used in human action recognition due to its view-invariant representation of pose structure, robustness to sensor noise \cite{zhang2012microsoft}, 
and efficiency in computation and storage \cite{li2018co,yan2017skeleton}. 
Recently, STGCN approaches have gained popularity by demonstrating superior performance in human activity understanding \cite{li2019actional, liu2020disentangling, das2021symmetric, pan2020spatio} and have become one of the state-of-the-art methods in the field of activity recognition. 
%Consequently, in this paper, we utilize skeleton-based activity recognition as an illustrative task to present the outcomes of our layer-wise analysis applied to STGCN.
As will be shown experimentally, our proposed layer-wise analysis of STGCNs helps us to (i) understand their generalization, (ii) detect bias toward learning any particular feature, (iii) evaluate model invariance to a set of functions, and (iv) assess robustness to perturbations to the input data. 
For example, in STGCN  for skeleton-based activity recognition \cite{yan2018spatial}, 
some layers may focus on learning the motion of specific body parts. 
Therefore, some models will not be suitable for new action classes where the motion is localized in other body parts. 

While layer-wise analysis of CNN  models \cite{bonet2021channelredundancy} and feature visualization methods \cite{simonyan2013deep, selvaraju2017grad} have been studied in the literature \cite{krizhevsky2012imagenet, he2016deep}, 
to the best of our knowledge, 
there exists \emph{no study} on the layer-wise explainability of the embeddings learned on spatiotemporal data using STGCNs. 
Moreover, layer-wise feature visualization techniques for STGCNs are also not available. 
%which hinders these STGCN networks to easily generalize to different tasks or in transfer learning.
In fact, most of the work on STGCN interpretation has studied only the final layer \cite{li2019actional, das2022gradient}. Extending the layer-wise analysis to STGCNs is not straightforward because of the varying lengths of the STGCN embeddings. This variability makes it difficult to find the similarity of embeddings of data points, such as action sequences with differing lengths, as the commonly employed similarity metrics  (e.g., cosine similarity or Euclidean distance)  are unsuitable for sequences of varying lengths.

Our first major contribution is a \emph{geometric} framework to characterize the data manifolds corresponding to each STGCN layer output. Our approach analyzes these manifolds by constructing a Non-Negative Kernel (NNK) DS-Graph \cite{shekkizhar2019graph} 
(\autoref{sec:NNK}), where nodes represent input sequences (actions) and distances between nodes are computed using dynamic time warping (DTW) \cite{muller2007dynamic} (\autoref{sec:DTW}).  This allows a distance to be computed between actions with different durations. 
We choose the NNK construction due to its robust performance in local estimation across different machine learning tasks \cite{shekkizhar2021model}. 
The benefits of the NNK construction will be demonstrated through a comparison with $k$-NN DS-Graph constructions in \autoref{sec:label_smoothness_ntu60}.
%showing how NNK graph successfully finds better neighbors over $k$-NN graph.
For the DS-Graph at each layer, we quantify the label smoothness as a way to track how the STGCN learns (\autoref{fig:summary}).  

Our approach has several important advantages: (1) the analysis is agnostic to the training procedure, architecture, or loss function used to train the model; (2) it allows for the comparison of features having different dimensions; (3) it can be applied to data that were not used for training (e.g., unseen actions or data in a transfer setting); (4) it allows us to observe how the layerwise representations are affected by external noise added to the input. 

%We did further studies in graph construction of the proposed geometric framework. In light of several drawbacks associated with $k$-nearest neighbor ($k$-NN) techniques, such as the arbitrary selection of parameters and the disregard for the relative positions of neighboring parameters, we opted to 

%We note that $k$-nearest neighbor ($k$-NN) \cite{peterson2009k}, which is a popular neighborhood construction technique, selects points in a neighborhood based on the only distance to the query point, without considering their relative positions, while also relying on ad hoc procedures to select parameter values of $k$. For this reason, we choose NNK to define neighborhoods and graphs for our manifold analysis. Unlike $k$-NN, which can be seen as thresholding approximation, NNK can be interpreted as a form of basis pursuit \cite{tropp2004topics}, which leads to better neighborhood construction with robust local estimation performance in several machine learning tasks \cite{shekkizhar2021model}. We also present a comparative analysis between $k$-NN and NNK graph construction showing how NNK graph successfully finds better neighbors over $k$-NN graph.

Our second major contribution is to extend our previous method, spatiotemporal graph GradCAM (STG-GradCAM) \cite{das2022gradient}, to perform layerwise visualization of the contributions of different Skeleton-Graph (S-Graph) nodes. To achieve this, we merge the class-specific gradient for a datapoint at each layer with the learned representations by that layer. This enables us to interpret individual layers within an STGCN network. The resulting layerwise STG-GradCAM (L-STG-GradCAM) allows us to visualize the importance of any node in any STGCN layer for the classification of a particular query class (action). 
This visualization helps confirm the results obtained through our analysis of the STGCN model using NNK-based geometric methods. It enhances the transparency of the model and deepens our comprehension of the representations learned at each layer. % This visualization serves to validate the findings from our NNK-based geometric analysis line layerwise interpretability of STGCN, enhancing model transparency and improving our understanding of the representations learned at each layer.

With our proposed data-driven label smoothness and layerwise visualization from L-STG-GradCAM, we can show that:
%\noindent Our contributions can be summarized as follows:
%    \item We develop Layerwise STG-GradCAM, a visualization tool for use at all layers of the STGCN network.
    % and provide insights into the working of an STGCN model by characterizing the input-output mapping induced by each successive layer of the model. 
(1) Initial layers learn low-level features corresponding to general human motion, while specific actions are recognized only in the later layers. 
(2) In a transfer task, the choice of which layers to leave unchanged and which layers to fine-tune can be informed by the changes in label smoothness for the target task on a network trained for the source task. 
(3) Experimentally, the label smoothness of an STGCN model over the layers as measured in the dataset graph is not affected significantly when Gaussian noise is added to the inputs, which justifies the observation that the model is robust to noise.

\section{Preliminaries}
\subsection{Skeleton graph and polynomial graph filters}
A skeleton graph (S-Graph) is a fixed undirected graph $\mathcal{G}_S = \{\mathcal{V}, \mathcal{E}, \hat{\Am}\}$ composed of a vertex set $\mathcal{V}$ of cardinality $|\mathcal{V}| = N$, an edge set $\mathcal{E}$ connecting vertices, and $\hat{\Am}$, a weighted adjacency matrix. 
$\hat{\Am}$ is a real symmetric $N \times N$ matrix, where $a_{i,j} \geq 0$ is the weight assigned to the edge connecting nodes $i$ and $j$. 
An STGCN layer (\autoref{sec:stgcn}) is a function of this adjacency matrix $\hat{\Am}$ and the identity matrix $\Id$ representing a self-loop. 
Specifically, STGCN uses the normalized adjacency matrix  
$\Am=\Dm^{-\frac{1}{2}}(\hat{\Am}+\Id)\Dm^{-\frac{1}{2}}$ where $(\Dm)_{ii}=\sum_ {j} (a_{i,j}+1)$. 
Intuitively, the elementary graph filter $\Am$ 
combines graph signals from adjacent nodes. Self-loops are added so that a node's own features are combined with those of its neighbors for learning. %The self-loops are added so that features associated with a node are combined along with its neighbor. 
%in each layer of applying the filter 
%and corresponds to  in the spectral domain of the adjacency matrix as a diagonal operator. 
%and corresponds to a diagonal filtering operator of the input in the eigenbasis of the adjacency. 

\begin{figure}[htbp]
\centering
\centering
    \includegraphics[width=.8\linewidth]{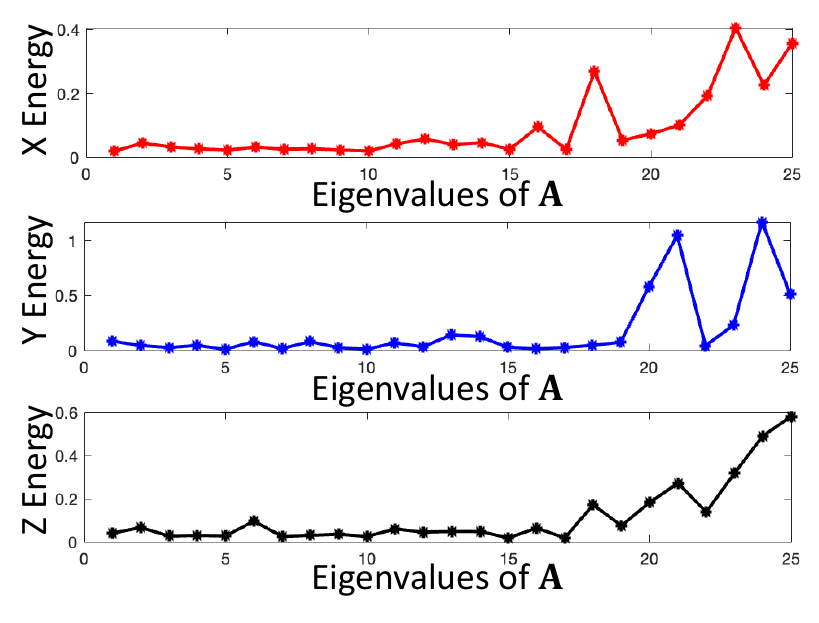}

   \caption{Energy graph spectrum of the human actions (NTURGB 120 \cite{liu2019ntu})  of the normalized adjacency matrix of the S-Graph ($\Am$). We use the graph spectrum of the adjacency matrix, as used in the STGCN, for ease of understanding. %However, a similar observation can be made using the graph Laplacian \cite{ortega2022introduction} as the eigenvectors for both cases are the same. 
   }
\label{fig:energyspectrum_freq_polyfilter}
\end{figure}

\autoref{fig:energyspectrum_freq_polyfilter} provides an example of how human motion is projected onto the eigenvectors of  $\Am$, leading to energy  
%of the normalized adjacency matrix used in STGCN for skeleton data. 
that is typically concentrated in 
the eigenvectors corresponding to the larger eigenvalues of $\Am$ (i.e., $\lambda_{17}$,$\ldots$, $\lambda_{25}$)\footnote{Note that the larger eigenvalues of $\Am$ correspond to the smaller eigenvalues of the  graph Laplacian $2\Id - \Am$. Thus, energy concentration in the higher eigenvalues of $\Am$ shows that typical human motion is smooth. }.
In each layer, we use simple filters of the form $\boldsymbol{x}_{out}=\mathcal{\Am} \boldsymbol{x}_{in} \Wm$, where $\Wm$ are trainable weights.
Applying these simple one-hop filters in multiple successive layers allows us to learn over multi-hop graph neighborhoods, analogous to what could be achieved with higher-order polynomials of the adjacency matrix,   
%This corresponds to having polynomial filters of the adjacency matrix, 
%which is equivalent to applying multiple layers of graph filters in terms of neural networks. 
%In other words, 
where an $l$-degree polynomial captures the data in a $l$-hop neighborhood. 
For example, in the human activity recognition task, the S-Graph has $25$ nodes and $24$ edges \cite{shahroudy2016ntu}. 
For this tree-structured graph, the maximum distance between two leaf nodes is $10$. Thus, a $10$-degree polynomial can capture information about the entire graph. 
This justifies using at most 10 layers of STGCN units in the STGCN network under consideration, where each layer is a function of $\Am$, i.e., a polynomial of degree $1$.

%define signal smoothness here

\subsection{Spatio Temporal Graph Convolutional Network}
\label{sec:stgcn}
 
%Each layer in an STGCN model combines both spatial and temporal graph convolutions.  
%for spatial information and temporal convolutions to capture the spatiotemporal features needed for solving a specific learning problem. 
%
STGCN for action recognition was first adopted in \cite{yan2018spatial}, where the spatial graph represented the intra-body 
connections of joints.  
%and where both spatial and temporal graph convolutions are used.  
While considering a spatiotemporal signal $\textbf{x}$, the input feature is represented as a $C\times N\times T$ tensor, where $C, N,$ and $T$ represent the number of channels, number of joints, and the temporal length of the activity sequence, respectively.  Graph convolutions are performed in two stages. First, a convolution is performed with a temporal kernel of size $(1 \times \tau)$.  Second, to capture the intra-joint variations, the resulting tensor is multiplied by the normalized adjacency matrix $\Am$ along the spatial axis. 
Denoting the input and output features of an STGCN layer as $\boldsymbol{x}_{in}$ and $\boldsymbol{x}_{out}$, the STGCN mapping is given by $\boldsymbol{x}_{out}=\mathcal{\Am} \boldsymbol{x}_{in} \Wm$, 
% \begin{equation}
% \label{eqn:stgcn}
%     \boldsymbol{x}_{out}=\mathcal{\Am} \boldsymbol{x}_{in} \Wm
% \end{equation}
where $\Wm$ represents trainable weight vectors corresponding to multiple input channels. Another matrix $\Qm$ is introduced to learn the edge weights of the graph.
%The mapping is further improved by allowing for the network to flexibly attend to its neighbors via a second weight matrix $\Qm$. 
Thus, each STGCN layer is implemented as follows: 
\begin{equation}
    \label{eqn:stgcn_part}
    \boldsymbol{x}_{out}=\sum_{j}^{}\Dm_{j}^{-\frac{1}{2}} (\Am_j \otimes \Qm) \Dm_{j}^{-\frac{1}{2}} \boldsymbol{x}_{in} \Wm_j,  
\end{equation}
where $\otimes$ denotes the Hadamard product.

\vspace{-20pt}

\subsection{Non-Negative Kernel(NNK) neighborhoods}
\label{sec:NNK}
% The traditional methods of defining a neighborhood, such as K-nearest neighbor (KNN) and $\epsilon$-neighborhood, solely depend on the distance to the query point, do not consider the relative positions of the neighbors, and require choosing hyperparameters, such as $k$ or $\epsilon$. For these reasons, we use non-negative kernel regression~(NNK) neighborhoods and graphs \cite{shekkizhar2020graph} for our manifold analysis. 
% Unlike KNN, which is a thresholding approximation, NNK can be viewed as a form of basis pursuit \cite{tropp2004topics} and results in better neighborhood construction with improved and robust local estimation performance in various machine learning tasks \cite{shekkizhar2021model, shekkizhar2021revisiting}.

We use non-negative kernel regression~(NNK) neighborhoods and graphs \cite{shekkizhar2020graph} for our manifold analysis because this results in better neighborhood construction with improved and robust local estimation performance in various machine learning tasks \cite{shekkizhar2021model, shekkizhar2021revisiting}.  The key advantage of NNK is its geometric interpretation for each neighborhood constructed. While in KNN points $\mathbf{x}_j$ and $\mathbf{x}_k$ are included in the neighborhood of a data point $\mathbf{x}_i$ solely based on their  
similarity to $\mathbf{x}_i$, i.e., $s(\mathbf{x}_i,\mathbf{x}_j)$ and $s(\mathbf{x}_i,\mathbf{x}_k)$, 
in NNK  this decision is made by also taking into account the metric $s(\mathbf{x}_j,\mathbf{x}_k)$. Consequently, $\mathbf{x}_j$ and $\mathbf{x}_k$ are both included in the NNK neighborhood only if they are not geometrically \emph{redundant}, the details are given in (\ref{sec:NNK}) .
%The obtained NNK neighborhoods can be described as a convex polytope approximation of the data point, determined by the local geometry of the data. This is particularly important for data that lies on a lower dimensional manifold in high dimensional vector space, a common scenario in feature embeddings in deep neural networks (DNNs).
NNK uses KNN as an initial step, with only a modest additional runtime requirement \cite{shekkizhar2020graph}. The computation can be accelerated using tools \cite{johnson2019billion} developed for KNN when dealing with large datasets. NNK requires kernels with a $[0, 1]$ range. In this work, we use the cosine similarity with the windowed aggregation in  \eqref{eqn:wdtw}. This kernel is applied to representations obtained after ReLU and satisfies the NNK definition requirement.

%The key advantage of NNK is that it has a geometric interpretation for each neighborhood constructed. 
%The obtained NNK neighborhoods can be described as a convex polytope approximation of the data point, determined by the local geometry of the data. This is particularly important for data that lies on a lower dimensional manifold in high dimensional vector space, a common scenario in feature embeddings in deep neural networks (DNNs).

\subsection{Dynamic Time Warping}
\label{sec:DTW}
%Dynamic Time Warping (DTW) is a versatile method for aligning time-dependent sequences, offering a superior alternative to Euclidean distance. 
While Euclidean distance permits only one-to-one point comparison, Dynamic Time Warping (DTW) (sec~ \ref{sec:NNK}) accommodates many-to-one comparisons, allowing precise alignment while considering temporal variations. %This feature makes DTW particularly effective for comparing arrays or time sequences of varying lengths. Essentially, DTW flexibly warps sequences to achieve optimal alignment. 
In our action recognition task, we work with action sequences of different durations, and our STGCN-based feature extraction retains temporal information. In this study, we employ DTW to measure the similarity between temporal features extracted by STGCN. DTW is computed using \ref{eqn:dtw}, where $\text{dtw}(i, j)$ represents the minimum warp distance between two time series of lengths $i$ and $j$. Each element in the accumulated matrix reflects the DTW distance between series $A_{1:i}$ and $B_{1:j}$.
 
  \begin{equation}
  \label{eqn:dtw}
  \begin{split}
        \text{DTW}(i,j) & =\text{dist}(a_i,b_j)+  \min(\text{DTW}(i-1,j), \\
        & \text{DTW}(i,j-1), \text{DTW}(i-1,j-1))
    \end{split}
 \end{equation}

\section{Proposed Geometric analysis of STGCN}
\label{sec:knn_labelsmoothness}

\subsection{Neighborhood analysis using dynamic DTW}

% \begin{itemize}
%     \item Use DTW metric to obtain k-nearest neighbors of an input's embedding at each STGCN layer
%     \item Observe the label smoothness via nearest neighbor label prediction.  
%     \item Observations can be made per action 
% \end{itemize}
Once we have a fully trained STGCN network, we construct an NNK DS-Graph using the representation generated by each layer of the STGCN model 
and refer to this graph as the NNK \textit{NNK Dataset Graph} $\mathcal{G}_D$. Note that each node corresponds to a data point in our \textit{NNK DS-Graph}, i.e., an action sequence represented by its features (learned by the STGCN). This differs from the S-Graph used in the STGCN model, which provides the original representation of an action sequence from which the features are extracted. After \textit{NNK DS-Graph} construction, we observe the smoothness of the class labels with respect to the graph, as shown in \autoref{fig:summary}. Graph smoothness or label smoothness in a graph represents the variation of the label of the neighboring node for each node in the DS-Graph. A DS-Graph has higher label smoothness when there is less variation in the labels of neighboring nodes. Our work uses label smoothness as a metric for assessing the representation of different layers within a network.

The main challenge with spatiotemporal action data is that each individual activity corresponds to a data sequence with a different temporal length. 
%For implementation, each sequence is zero-padded to make an equal-length input. But, the zero values resulting from padding do not correspond to the actual input action sequence. 
%Therefore, a Euclidean distance or cosine similarity is not applicable to find similarity between two sequences. This makes it difficult to compute the distance between two input sequences or actions. 
To address this issue, we develop a  DTW-based distance metric to find the similarity between the representations (\autoref{sec:DTW}).  Computation of this \textit{window-based DTW} distance metric \textit{w-DTW}  involves the following steps. 
\begin{itemize}%[leftmargin=*]
    \item Consider two sequences $\boldsymbol{s}_i$ and $\boldsymbol{s}_j$ divided temporally into $m$ windows. The dimension of $\boldsymbol{s}_i$ and $\boldsymbol{s}_j$ is  $N\times T_i$ and $N\times T_j$ respectively. Here $N$ denotes the number of spatial joints and $T_i$ denotes the temporal length of the $i^{th}$ sequence.
    \item $\boldsymbol{s}_i^{w}$ denotes the $w$-th window of the sequence, then the distance between two sequences is computed as follows.
    \begin{equation}
    \text{wDTW} (\boldsymbol{s}_i , \boldsymbol{s}_j) = \sum_{w=1}^{m} \alpha_w  \text{DTW}(\boldsymbol{s}_{i}^{w},\boldsymbol{s}_{j}^{w})
    \label{eqn:wdtw}    
    \end{equation}
    $\alpha_w$ is the weight to the $w$-th window, $\sum_{w=1}^{m}\alpha_w=1$. 
    \item The weights are chosen such that they decrease along the temporal axis based on the length statistic of all the sequences in the dataset i.e., the number of samples that have non-zero padding in a particular temporal window.
    % to take into account the effect of the padding which does not contribute to the real feature. %. As the sequences are zero-padded at the end, we give a smaller weight to the later windows.
\end{itemize}

% This method not only helps us to align the sequences and find the distance, but the \textit{wDTW} also makes the processing faster. Thus, using the proposed wDTW-based $k$-NN, we observe the data points after each layer of STGCN via a graph representation and draw insights about its property using label smoothness.

%% We need to say why label smoothness and connect to the theorem below
%The goal of our NNK-based analysis is to demonstrate that, 

While STGCN involves complex mappings, the transformations they induce and the corresponding structure of each representation space can be studied using a graph constructed on the embedded features. 
Consider an STGCN model and a spatiotemporal dataset. At each layer, all sequences in the dataset can be represented using the \textit{NNK Dataset Graph}. In this graph, each node corresponds to a sequence, and the action labels are treated as 'signals' or attributes associated with these nodes, as illustrated in \autoref{fig:summary}. At the output of each layer, each input sequence is mapped to new values (in some other feature space). Thus, we can associate a new NNK DS-graph to the same set of data points (with the same signal, i.e., label).
Thus, instead of directly working with the high dimensional features or the model's overparameterized space, the focus is on the relative positions of the feature embeddings obtained in STGCN layers. This allows us to characterize the geometry of the manifold spaces encoded by an STGCN and to develop a quantitative understanding of the model.

\begin{figure*}[htbp]
\begin{center}
\includegraphics[width=1\linewidth]{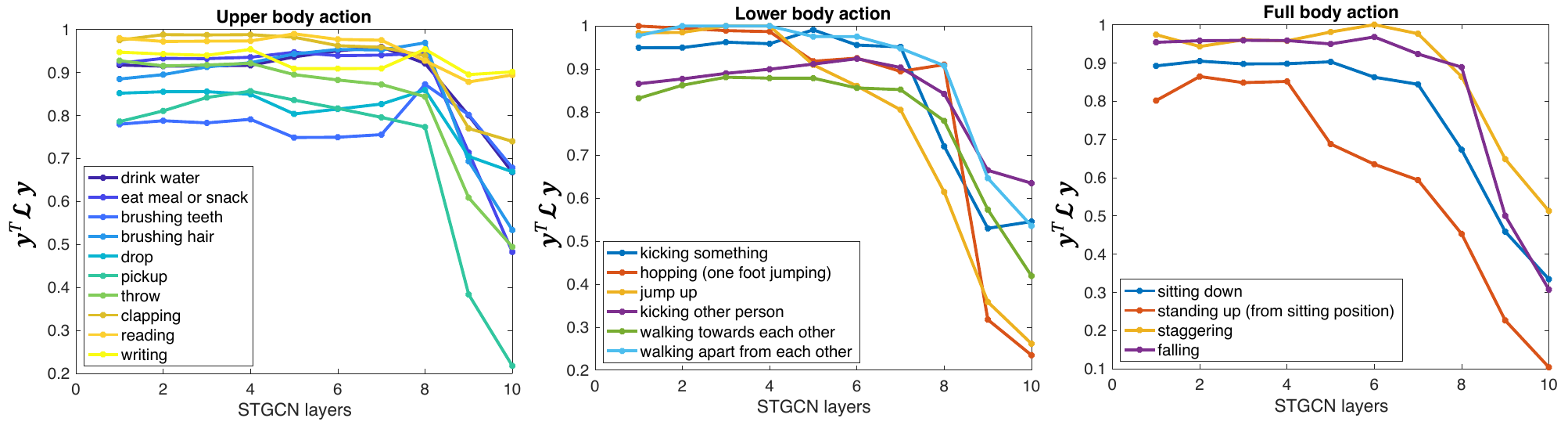} 
\end{center}
   \caption{Smoothness of labels on the manifold induced by the STGCN layer mappings in a trained model. As the label smoothness increases, the Laplacian quadratic ($ \mathbf{y}^\top\mathcal{L}\mathbf{y}$) decreases. Intuitively, a lower value of $ \mathbf{y}^\top\mathcal{L}\mathbf{y}$ corresponds to the features belonging to a particular class having graph neighbors from the same class.We divide the actions in NTU-RGB60 into three super-classes~(Upper body (\textbf{Left}), Lower body (\textbf{Middle}), Full body (\textbf{Right})) and present smoothness with respect to each action in the grouping. We emphasize that, though the smoothness is displayed per class, the \textit{NNK Dataset Graph} is constructed using the features corresponding to all input action data points. We observe that the model follows a similar trend, where the smoothness of labels is flat in the initial layers (indicative of no class-specific learning) and increases in value in the later layers (corresponding to discriminative learning). Outliers exist to this trend~(e.g., in upper body group \emph{drop, brushing}) where the smoothness decreases in intermediate layers. This may imply that the representations for these actions are affected by features from other actions to accommodate for learning other classes.}
\label{fig:nturgb60_actwise_smoothness}
\end{figure*}

We now present a theoretical result (Theorem \ref{thm:label_smoothness}) relating the respective label smoothnesses of the input and output features of a single layer in a neural network to that of its complexity measured by the $\ell_2$-norm \cite{gunasekar2018implicit, ongie2019function}. 
The proof for the theorem is provided in the supplementary materials (\autoref{sec:proof_thoerem_smoothness}).

\begin{definition}[Label smoothness] 
Given a graph represented by its Laplacian $\mathcal{L}$ and a label signal $\mathbf{y}$ on the graph, the Laplacian quadratic $ \mathbf{y}^\top\mathcal{L}\mathbf{y}$ captures the smoothness of the label on the graph \cite{zhu2003semi, ortega2022introduction}. 
Note that smaller values of $\mathbf{y}^\top\mathcal{L}\mathbf{y}$ correspond to smoother signals. In other words, an increase in the label similarity of the connected nodes is commensurate with a decrease in $\mathbf{y}^\top\mathcal{L}\mathbf{y}$.
\end{definition}
\begin{theorem}
\label{thm:label_smoothness}
    Consider the features corresponding to the input and output of a layer in a neural network denoted by $\boldsymbol{x}_{out} = \phi(\mathbf{W}\boldsymbol{x}_{in})$ where $\phi(\boldsymbol{x})$ is a slope restricted nonlinearity applied along each dimension of $\boldsymbol{x}$. Let us suppose that the smoothness of the labels $\boldsymbol{y}$ in the feature space is proportional to the smoothness of the data $\boldsymbol{x}$. Then,
    \begin{align}
        \mathbf{y}^\top\mathcal{L}_{out}\mathbf{y} \leq c\;||\mathbf{W}||_{2}^2 \;\; \mathbf{y}^\top\mathcal{L}_{in}\mathbf{y}
    \end{align}
    where $\mathcal{L}$ corresponds to the graph laplacian  obtained using NNK in the feature space. Note that $c > 0$ depends only on constants related to data smoothness and the slope of the non-linearity. 
\end{theorem}

\begin{remark}
    Theorem \ref{thm:label_smoothness} states that the change in label smoothness between the input and output spaces of a network layer is indicative of the complexity of the mapping induced by that layer, i.e., a big change in label smoothness corresponds to a larger transformation of the features space. 
\end{remark}
\begin{remark}
 Theorem \ref{thm:label_smoothness} does not make any assumption on the model architecture and makes an assumption about the relationship between the respective smoothness of the data and the labels. The slope restriction on the nonlinearity is satisfied by activation functions used often in practice. For example, the ReLU function is slope restricted between $0$ and $1$ \cite{fazlyab2020safety,fazlyab2019efficient}.
\end{remark}

The idea of characterizing intermediate representations using graphs was previously studied in \cite{gripon2018inside, lassance2021representing}. However, these works were limited to images and did not study spatiotemporal data. To the best of our knowledge, our work presents the first methodology for use with structured input sequences for analysis and understanding of STGCN networks. 

Our method uses NNK for analysis similar to \cite{bonet2021channelredundancy}. However, unlike other approaches, our work focuses on the geometry of the feature manifold induced by the STGCN layer using the NNK  graphs constructed. %Closer to our approach, \cite{shekkizhar2021model, bonet2021channelredundancy} study the geometry of the manifold induced by convolutional networks on images using a refinement of the $k$-NN graph \cite{shekkizhar2020graph}. It is not clear how the graph construction used in these related works can be adapted to spatiotemporal data and is left as future work.

% We find that up to a certain layer from the beginning, the network learns general human motion and then focuses on understanding specific actions. A similar trend is also found in the progression of the node importance in the network computed by the Layerwise STG-Grad CAM discussed later. This observation motivated us to analyze the transfer performance of the model in a new dataset NTU-RGB120 \cite{liu2019ntu}. We find a similar trend in the label smoothness over the layers of STGCN for NTU-RGB120 as NTU-RGB60 \cite{shahroudy2016ntu}. % well stating that the first few layers focus on overall human motion understanding while the last few layers focus on understanding specific action.
% The label smoothness also provides information about which layers we need to fine-tune to adapt to the new dataset. This reduces complexity and time while using STGCN in a new dataset. Lastly, we evaluate the performance of the noisy samples and show that STGCN is very robust with Gaussian noise. In fact, it helps to achieve better label smoothness by acting as a regularizer to the model.

\subsection{Layerwise (L) STG-GradCAM}
As in regular convolutional layers, unlike fully-connected layers, spatiotemporal graph convolution layers retain localized information both in the spatial and temporal axis. 
\cite{das2022gradient} proposed STG-GradCAM for visualizing the importance of the nodes in the spatiotemporal skeleton graph for a particular action. %This approach used the gradient corresponding to the target action flowing to the final convolutional layer \cite{selvaraju2017grad}.as a function of the previous layer's output and the spatiotemporal graph adjacency matrix. This helped interpret the spatiotemporal importance of different graph nodes for specific classification tasks. 
However, that work only used the last STGCN layer to provide an interpretation. In this paper, we extended STG-GradCAM to L-STG-GradCAM for use with all layers of an STGCN model.
% We know that the last STGCN layer contains high-level spatiotemporal information associated with an action \cite{das2022gradient}, however, the features of the rest of the layers are not studied in the case of STGCN. 
% The neurons in these layers are responsible for the class-specific spatiotemporal importance of the skeleton joints in the activity sequence data. % Repetitive

The gradient information flowing into each STGCN layer is used in our proposed L-STG-GradCAM to compute the importance of each neuron for a particular class prediction and to determine whether the intermediate layers are learning something meaningful. 
% These gradients carry the class-specific information to the intermediate layers. 
We use the gradients as the weight of the representations at each layer. The outcome of L-STG-GradCAM helps us to understand which part of the data in each layer contributes to the final decision. 
Let the $k^{th}$ graph convolutional feature map at layer $\ell$ be defined as: $ \Fm_k^\ell(\Xm,\Am)=\sigma (\Tilde{\Am}  \Fm^{\ell-1} (\Xm,\Am) \Wm_k^\ell)$.
% \begin{equation}
%     \label{eqn:grad_cam_im}
%     \Fm_k^\ell(\Xm,\Am)=\sigma (\Tilde{\Am}  \Fm^{\ell-1} (\Xm,\Am) \Wm_k^\ell)
% \end{equation}
Here, the $k^{th}$ feature at the $\ell^{th}$ layer is denoted by $\Fm^\ell_{k,n,t}$ for node $n$ and time $t$. Then, L-STG-GradCAM ’s label-specific weights for class $c$ at layer $l$ and for feature $k$ are calculated by: 
\begin{equation}
    \label{eqn:grad_cam_weight}
    \boldsymbol{\beta}_{k}^{c,\ell}=\frac{1}{NT} \sum_{n=1}^{N}\sum_{t=1}^{T}  \frac{\delta y^c}{\delta \Fm^{\ell}_{k,n,t}}.
\end{equation}
Then, we can compute the importance of the nodes in a specific layer $\ell$ using: 
\begin{equation}
    \label{eqn:grad_cam_heatmap}
    H_{ST}^{c,\ell}=\mathrm{ReLU} \left( \sum_k \boldsymbol{\beta}_{k}^{c,\ell} \Fm ^\ell_{k} \right).
\end{equation}

L-STG-GradCAM  enables us to visualize the class-specific spatiotemporal importance 
(\autoref{fig:Lstggradcam_kick}) of the representation for any layer $\ell$ of the network. The code is available \href{https://github.com/daspraty/stg-gradcam.git}{online}.  %This helps us to understand the importance of different nodes in a network as a function of spatial and temporal axis for a specific task (Fig. \ref{fig:Lstggradcam_kick}). 
% In other words, which part of the feature map in each layer is contributing to predicting a specific label for specific action? 

\section{Results}

\subsection{Experiment setting}
\label{sec:expsetting}

% \paragraph{Network Architecture:} 

\textit{a. Network Architecture:} %First, input skeletons are fed to a batch normalization layer to keep the scale of input data consistent on different joints. 
The STGCN model used for human action recognition by \cite{yan2018spatial} comprises $10$ STGCN layers implemented as in \autoref{sec:stgcn}. The first four layers have $64$ output channels, the next three layers have $128$, and the last three have $256$. Afterward, a global pooling layer with a softmax is used as a classifier. The model is trained using cross-entropy loss with batch SGD for $100$ epochs.

% \paragraph{Datasets:}
\textit{b. Datasets:}
\textit{NTU-RGB60}: We use the STGCN model described above and introduced in \cite{yan2018spatial} and train the model in cross-subject (x-sub) settings on NTU-RGB60 \cite{shahroudy2016ntu} dataset. This dataset contains $56,000$ action clips corresponding to $60$ action classes performed by $40$ subjects, e.g., Throw, Kick). The dataset includes annotated 3D joint locations (X, Y, Z) of $25$ joints. \textit{NTU-RGB120}: NTU-RGB120 \cite{liu2019ntu} extends NTU-RGB60 with an additional $57,367$ skeleton sequences over $60$ extra action classes, from $106$ distinct subjects.

\subsection{Label smoothness computed from the features}
\label{sec:label_smoothness_ntu60}
To provide insights into the representations of the intermediate layers of the STGCN network, we make use of our geometric analysis of the representation using the DTW-based NNK method described in \autoref{sec:knn_labelsmoothness}. 
\autoref{fig:summary} (right) shows the label smoothness over the layers of STGCN for different sets of the upper body, lower body, and full body actions (refer to \autoref{tab:ntu60_actioncategory}). We see a sudden fall in the Laplacian quadratic after layer $8$, while the slope is small before that layer. This implies that the early layers have features that are mostly not class-specific. In contrast, the smoothness improves (corresponds to a decrease in the value of $ \mathbf{y}^\top\mathcal{L}_{out}\mathbf{y}$) using the representations after layer $8$ consistently across all input actions. Following 
\autoref{thm:label_smoothness}, we can state that the large change in the label smoothness from layer $8$ to $9$ corresponds to a larger transformation (equivalent to the functional norm of the layer is large) in the input-output mapping of this layer. Our earlier visualization using L-STG-GradCAM validates this analysis visually.
% as presented in section~\ref{sec:LSTG-GradCAM}. 
%, which interestingly also exhibits a significant change between layer 7 and layer 8.
\autoref{fig:nturgb60_actwise_smoothness} presents action-wise label smoothness over the layers of STGCN. 
This figure helps us better understand which actions are learned over the layer of STGCN. 
For example, 
in \autoref{fig:nturgb60_actwise_smoothness} (left), action \textit{reading} and \textit{writing} are poorly learned. The value of $ \mathbf{y}^\top\mathcal{L}_{out}\mathbf{y}$ in the plot is not monotonically decreasing for all the actions. For example, for the action \textit{drop}, the label smoothness decreases in the middle and increases again at the end. The possible reason behind this pattern is that the network tries to accommodate other actions and again learns all the actions gradually before the last layer.

\begin{figure}[H]
\centering
\centering
    \includegraphics[width=.6\linewidth]{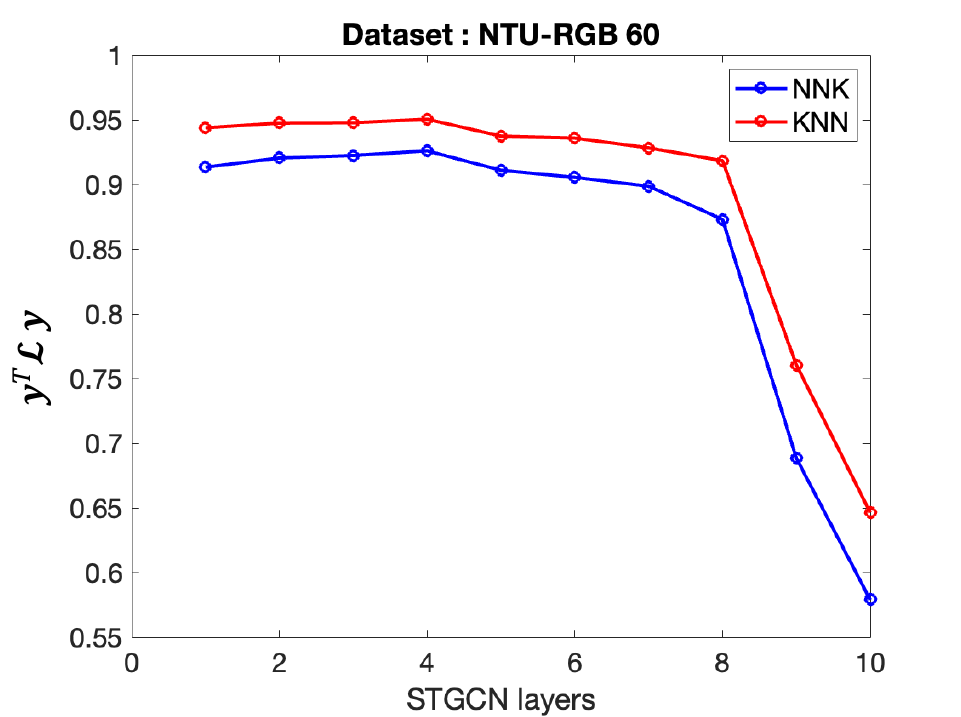}
   \caption{Label smoothness of STGCN for different DS-graph construction methods (blue)-NNK, (red)-$k$-NN.} %As the label smoothness increases the Laplacian quadratic ($ \mathbf{y}^\top\mathcal{L}\mathbf{y}$) decreases.}
\label{fig:NNKvsKNN}
\end{figure}

\textbf{Comparison between NNK and $k$-NN}: \autoref{fig:NNKvsKNN} shows the effect on label smoothness for different choices of graph construction methods like NNK and $k$-NN. Higher label smoothness (small value of $ \mathbf{y}^\top\mathcal{L}_{out}\mathbf{y}$) represents better the construction of the graph, reducing the prediction error at each layer. %Better graph construction implies better choices of the neighbors and their corresponding weights.  
NNK clearly performs better than $k$-NN in choosing the right neighbors and their corresponding weights.

% \vspace{-5pt}

\begin{figure*}[htbp]
\centering
\includegraphics[width=0.9\linewidth]{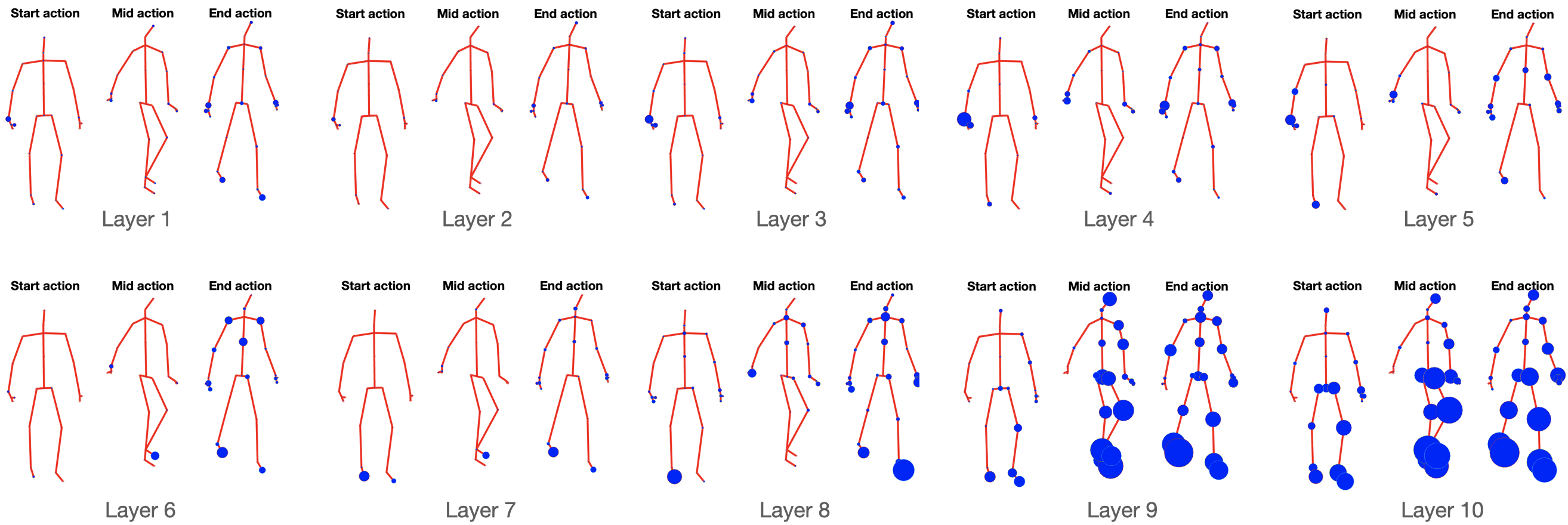}  \caption{\label{fig:Lstggradcam_kick} L-STG-GradCAM visualization of spatiotemporal node importance for action class \emph{Kick} of a trained STGCN network used in experiments. The size of the blue bubble denotes the relative importance of the node in a layer for prediction by the final softmax classifier and is scaled to have values in $[0, 1]$ at each layer. The node importance values are normalized across layers to have a clear comparison among the layers. We observe that the localization of the action as observed using the L-STG-GradCAM is evident only in later layers while initial layers have no class-specific influence. The visualizations allow for transparency in an otherwise black-box model to explain any class prediction. Our approach is applicable to any STGCN model and is not affected by the model size, optimization strategy, or dataset used for training. }
\end{figure*}

\subsection{L-STG-GradCAM  visualization}
\label{sec:LSTG-GradCAM}
%We consider a human action classification task using skeleton data to evaluate the performance of STG-Grad CAM extended for each layer. The $10$-layer STGCN model \cite{yan2018spatial} is used for evaluation. 
The STGCN model in \autoref{sec:expsetting} with NTU-RGB60 achieves $82.1\%$ accuracy on NTU-RGB60 \textit{xsub} setting \cite{yan2018spatial}. We use  \eqref{eqn:grad_cam_heatmap} to generate a class-specific skeleton joint-time importance map using all the data points corresponding to a given class. 
\autoref{fig:Lstggradcam_kick} shows the layerwise variation of joint importance for the action \textit{'Kick'} for three time-slices. The node's size in the S-graph denotes the degree of importance of the body joint at that time point for the final prediction. For the action \emph{Kick}, which mostly involves lower body parts, we notice in the figure that the initial layers (up to layer $8$) have very weak, if any, GradCAM localization corresponding to the action. In contrast, the last three STGCN layers show explicit node importance \emph{heatmap} where the leg and the back joints are relatively more active, indicative of the action. We present additional examples in the supplementary (\autoref{fig:stggradcam_throw}, \autoref{fig:stggradcam_sitdown}) corresponding to an upper-body action (\textit{Throw}) and a full body action (\textit{Sitting down}). In both cases, we find a similar trend where the STGCN graph filters learned in the initial layers capture general human motion, focusing on all the nodes in the S-graph and having class-specific node importance only in a few final layers of the network.

\subsection{Effect of noise in the data}

We analyze the robustness of the STGCN network in the presence of noise in the data. In our experiments, we add noise at various peak signal-to-noise ratio (PSNR) levels to a set of actions and compare the label smoothness over the layers concerning the original signal.

\begin{figure*}[htbp]
\begin{center}
\includegraphics[width=.9\linewidth]{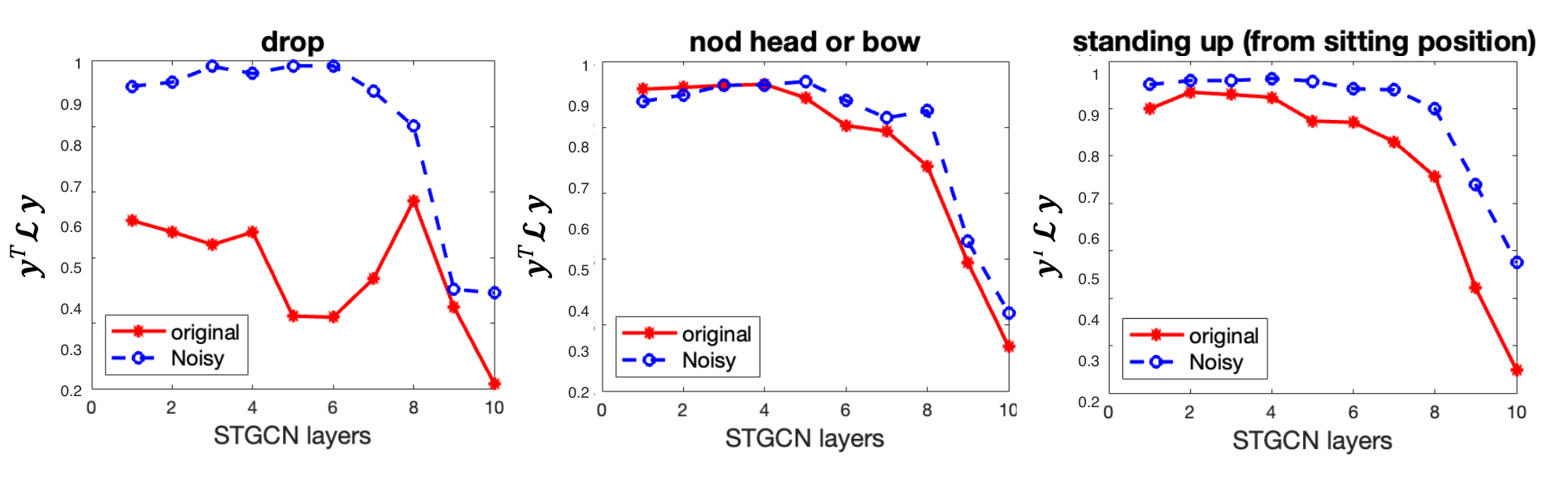}   
\end{center}
\caption{Impact of noise added to the input sequence on the label smoothness observed using the corresponding features obtained with the model. The Laplacian quadratic ($ \mathbf{y}^\top\mathcal{L}_{out}\mathbf{y}$) decreases as the label smoothness increases. We show the impact of noise on one action per super-class grouping (Upper, Lower, and Full Body). We observe that in actions 
where the Laplacian quadratic had a steady (non-increasing) trend, it remained mostly unaffected by adding noise to the input action sequence. However, actions where the label smoothness decreased before increasing were affected by noise. This implies that the features learned in the early layers for these actions are not robust, and adding noise allows us to see the modified manifold induced in these layers. }
\label{fig:noisysamples}
\end{figure*}

%In our experiments, we add noise at various peak signal-to-noise ratio (PSNR) levels to a set of 12 actions (please refer Fig.~\ref{fig:noisy12}) in the dataset and compare the label smoothness over the layers concerning the original signal. %These actions are \textit{'drop','pickup','sitting down','standing up (from sitting position)','take off a hat or cap','cheer up','hopping (one foot jumping)','jump up','nod head or bow','staggering','falling','touch neck (neckache)','handshaking'} .These actions are chosen based on the good performance observed in the label smoothness.
A popular approach to incorporate noise into the spatiotemporal data is to add additive white Gaussian noise to the measurement \cite{kao2019graph}.  \autoref{fig:noisysamples} shows label smoothness for three actions \textit{drop}, \textit{hop}, and \textit{standing up (from a sitting position)}. It is clear from these examples that the overall performance degraded slightly, while the smoothness of the labels through successive layers of the network is better than the original signal.% Therefore, we postulate that the noise added acts as a regularizer for the induced manifold of the model and helps the network gradually improve.
  Specifically for the action \textit{drop}, as we discussed in \autoref{sec:label_smoothness_ntu60}, the label smoothness degraded in the middle of the network and recovered at the end. However, for the noisy signal, we notice a more stable, non-increasing pattern as in other actions. The accuracy of the STGCN network on this partially noisy dataset is $80.2\%$. Hence, the network is robust to this additive white Gaussian noise.
%Another approach is to add frequency selective noise to the data.As we have seen in fig. \ref{fig:energyspectrum_freq_polyfilter}, most of the human motion lies in the frequency range $\lambda_{17},\lambda_18,\lambda_19,...,\lambda_24,\lambda_25\}$. We attack the STGCN network by adding noise to $\lambda_1 and \lambda_{15}$

% \vspace{-0.4cm}

\subsection{Transfer performance}

So far, we see that the first few network layers focus on understanding general human motion, which is needed before learning the specific task. Therefore, the hypothesis is that the network should exhibit similar behavior in the layerwise representations for a similar human activity dataset. To explore the area of adapting the pre-trained STGCN model to a new dataset and analyze the transfer performance of STGCN, we use the new 60 actions (61-120) in NTU-RGB120 \cite{liu2019ntu} dataset. In the rest of the paper, we refer to these new actions 61-120 as NTU-RGB61-120. 

\begin{figure}[htbp]
% \centering
\begin{center}
\includegraphics[width=1\linewidth]{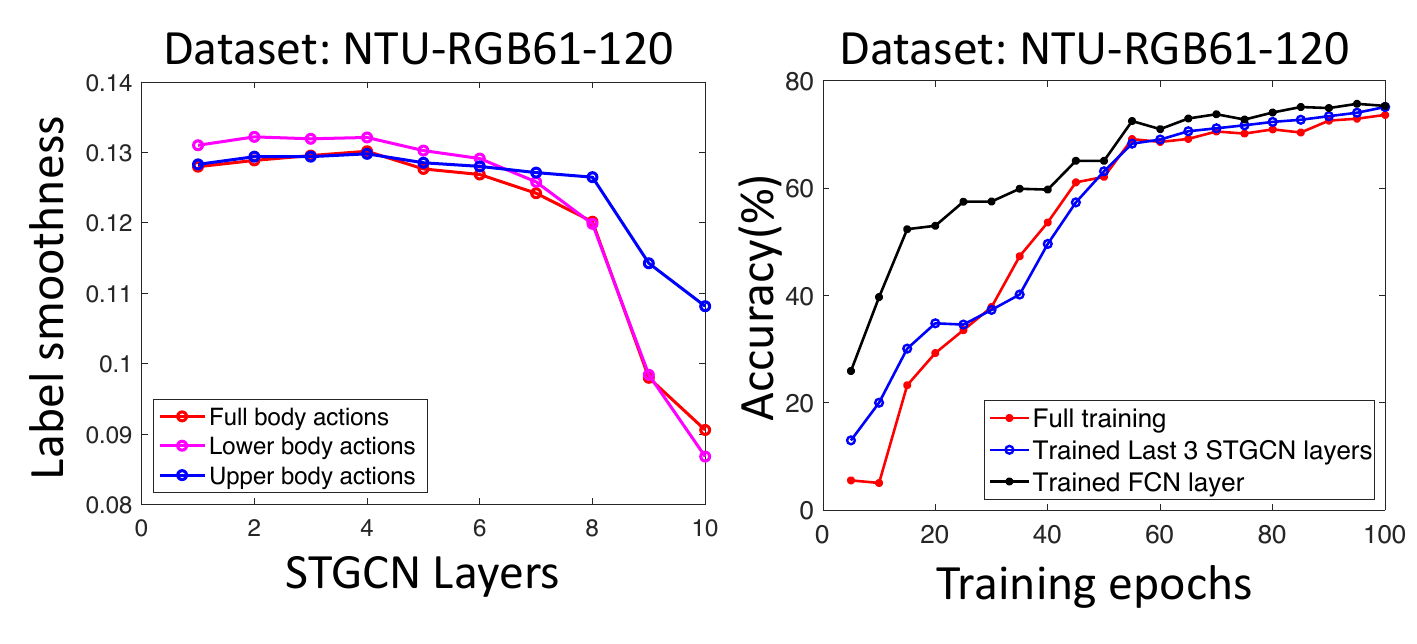}   
\end{center}
\caption{ \textbf{Left:} Label smoothness of unseen action classes (NTU-RGB61-120) using a model trained on NTU-RGB60. We present results averaged over each super-class~(Upper, Lower, and Full body). We see that the model embedding allows the features corresponding to new action sequences to be separable. Further, $\mathbf{y}^\top\mathcal{L}\mathbf{y}$ follows a similar non-increasing trend as in the case of the NTU-RGB60 in a much smaller range of scale. This implies that the features learned by the model can be used for the novel classes, and model transfer can be done with simple fine-tuning.  \textbf{Right:} Classification accuracy on NTU-RGB61-120 test-set using a $10$-layer STGCN network. Performance comparison between a model trained from scratch and one obtained with transfer learning by fine-tuning a model trained on NTU-RGB60. We can see the effectiveness of model transfer, which was predicted by our label smoothness analysis.}
\label{fig:nturgb120}
\end{figure}

We first analyze the label smoothness of the NTU-RGB61-120 dataset on the pre-trained STGCN model trained on the NTU-RGB60 dataset. \autoref{fig:nturgb120}~(Left) shows the label smoothness over the successive layers of the network. We divided NTU-RGB61-120 into three sets, lower body, upper body, and full body actions, depending on the involvement of the body joints (\autoref{tab:ntu120_actioncategory}) We notice a similar pattern as we observed for NTU-RGB60. There is a big jump in the label smoothness after layer $8$, while the slope changes slowly before that. Therefore, although the network is not trained on NTU-RGB61-120, it shows similar behavior, proving our hypothesis. The overall accuracy of the network is $9\%$, which states the need for fine-tuning and it achieves $~78\%$ accuracy after fine-tuning.

\autoref{fig:nturgb120}~(Right) shows the validation accuracy (validation loss in supplementary~\autoref{fig:ntu120_valloss} )of the STGCN network with respect to the training epochs. In the case of transfer learning, we can fine-tune the last few layers depending on the performance or the availability of the data.  We consider 3 cases of fine-tuning varying the number of layers, such as 1. training only the FCN layer, 2. training the last 3 STGCN layers, including the FCN layer, and 3. training the whole network. Interestingly, we see that in this case, only fine-tuning the FCN layer (case 1) provides good performance. This means that STGCN captures a good representation of these human motions. If we have a dataset where the actions are very different than the trained dataset, we can fine-tune more layers depending on the availability of the data.

% \begin{figure}[htbp]
% \centering
% \begin{subfigure}[t]{0.36\textwidth}
%    \includegraphics[width=.8\linewidth]{images/nnk_ntu120_actionwise.pdf}
% % \label{fig:nturgb120_labelsmoothness}
% \end{subfigure}
% ~
% \begin{subfigure}[t]{0.34\textwidth}
% \begin{center}
%    \includegraphics[width=.75\linewidth]{images/nturgb120_acc_3instances.pdf}
% \end{center}
% % \label{fig:nturgb120_acc}
% \end{subfigure}
% \caption{ \textbf{Top:} Label smoothness of unseen action classes (NTU-RGB61-120) using a model trained on NTU-RGB60. We present results averaged over each super-classes~(Upper, Lower, and Full body). We see that the model is able to embed the features corresponding to new action sequences such that they are separable. Further, $\mathbf{y}^\top\mathcal{L}\mathbf{y}$ follows a similar non-increasing trend as in the case of the NTU-RGB60 in a much smaller range of scale. This implies that the features learned by the model can be used for the novel classes and model transfer can be done with simple fine-tuning.  \textbf{Bottom:} Classification accuracy on NTU-RGB61-120 test-set using a $10$-layer STGCN network. Here, we show the performance achieved by the model when training from scratch, along with that obtained with transfer learning by fine-tuning a model trained on NTU-RGB60. We see that the model is able to transfer effectively, as predicted by our label smoothness analysis.}
% \label{fig:nturgb120}
% \end{figure}

\section{Conclusion}
We present a data-driven approach for understanding STGCN models using windowed-DTW distance-based NNK graphs.
% We focused on a skeleton-based activity recognition task, but this is a generic method to analyze any DNN dealing with spatiotemporal data even with varying temporal length. 
Analyzing the label smoothness of the successive layers on the \textit{NNK Dataset Graph}, we show that the initial layers focus on general human motion, and features for individual action recognition are learned by the model only in the later layers. We also present a comparison between graph construction methods, showing the superiority of the NNK graph over the $k$-NN graph. To validate our insights from label smoothness, we introduce an L-STG-GradCAM method to visualize the importance of different nodes at each layer for predicting the action. We then present our analysis of label smoothness and its impact on the transfer performance of a trained STGCN model to unseen action classes. Finally, we present an analysis of the robustness of the features at each layer of an STGCN in the presence of input Gaussian noise. We show that the added noise %to each skeleton joint 
does not affect the label smoothness of several action classes. %  and acts as a regularizer on the embedding manifold induced by the model. A detailed discussion on the robustness of the STGCN model in the presence of various noise models is left for future work.

% \pagebreak
\clearpage
\newpage
{\small
\bibliographystyle{IEEEtran}
\bibliography{main}
}

\clearpage
\newpage

\onecolumn

\section{Supplementary Material}
\subsection{Non-Negative Kernel Regression (NNK) neighborhoods}
\label{sec:NNK}
The traditional methods of defining a neighborhood, such as K-nearest neighbor (KNN) and $\epsilon$-neighborhood, solely depend on the distance to the query point and do not consider relative positions. Further, these methods also require an arbitrary selection of parameters, such as $k$ or $\epsilon$. For these reasons, we make use of non-negative kernel regression~(NNK) \cite{shekkizhar2020graph} to define  neighborhoods and graphs for our manifold analysis. 
Unlike KNN, which is a thresholding approximation, NNK can be viewed as a form of basis pursuit \cite{tropp2004topics} and results in better neighborhood construction with improved and robust local estimation performance in various machine learning tasks \cite{shekkizhar2021model, shekkizhar2021revisiting}.
The key advantage of NNK is that it has a geometric interpretation for each neighborhood constructed. While in KNN points $\mathbf{x}_j$ and $\mathbf{x}_k$ are included in the neighborhood of a data point $\mathbf{x}_i$ solely based on their  
metric to $\mathbf{x}_i$, i.e., $s(\mathbf{x}_i,\mathbf{x}_j)$ and $s(\mathbf{x}_i,\mathbf{x}_k)$, 
in NNK  this decision is made by also taking into account the metric $s(\mathbf{x}_j,\mathbf{x}_k)$. Consequently, $\mathbf{x}_j$ and $\mathbf{x}_k$ are both included in the NNK neighborhood only if they are not geometrically \emph{redundant}.
The obtained NNK neighborhoods can be described as a convex polytope approximation of the data point, determined by the local geometry of the data. This is particularly important for data that lies on a lower dimensional manifold in high dimensional vector space, a common scenario in feature embeddings in deep neural networks (DNNs).
NNK uses KNN as an initial step, with only a modest additional runtime requirement \cite{shekkizhar2020graph}. The computation can be accelerated using tools \cite{johnson2019billion} developed for KNN when dealing with large datasets. NNK requires kernels with a range in $[0, 1]$. In this work, we make use of cosine similarity with the windowed aggregation as in equation (\ref{eqn:wdtw}). The kernel is applied on representations obtained after ReLU and  hence satisfies the NNK definition requirement.

\subsection{Proof of Theorem \ref{thm:label_smoothness}}
\label{sec:proof_thoerem_smoothness}
\begin{proof}
    Let $\mathcal{N}^{(i)}_{in}, \mathcal{N}^{(i)}_{out}$ be the set of NNK neighbors of data point $\mathbf{x}^{(i)}$ in the input-output feature spaces. 
    \begin{align}
        \mathbf{y}^\top\mathcal{L}_{out}\mathbf{y} = \sum_i \sum_{j \in \mathcal{N}^{(i)}_{out}} \theta_{ij} | y^{(i)} - y^{(j)} |^2
    \end{align}
    Now, by assumption, the smoothness of the labels is proportional to the data smoothness. Thus,
    \begin{align}
        \mathbf{y}^\top\mathcal{L}_{out}\mathbf{y} = \sum_i \sum_{j \in \mathcal{N}^{(i)}_{out}} c_{out} \; \theta_{ij} ||\mathbf{x}^{(i)}_{out} - \mathbf{x}^{(j)}_{out}||^2
    \end{align}
    where $c_{out} > 0$ is the proportionality constant.

    $\mathcal{N}^{(i)}_{out}$ corresponds to the optimal NNK neighbors, and therefore any other  neighbor set will have a larger value and label smoothness, i.e.,
    \begin{align}
        \mathbf{y}^\top\mathcal{L}_{out}\mathbf{y} &\leq c_{out} \;\sum_i \sum_{j \in \mathcal{N}^{(i)}_{in}} \theta_{ij}||\mathbf{x}^{(i)}_{out} - \mathbf{x}^{(j)}_{out}||^2 \nonumber \\
        & = c_{out} \; \sum_i \sum_{j \in \mathcal{N}^{(i)}_{in}} \theta_{ij}||\phi(\mathbf{W}\mathbf{x}^{(i)}_{in}) - \phi(\mathbf{W}\mathbf{x}^{(j)}_{in})||^2 \nonumber \\
        &\leq c_{out}\; \sum_i \sum_{j \in \mathcal{N}^{(i)}_{in}} \beta \theta_{ij}\; ||\mathbf{W}\mathbf{x}^{(i)}_{in} - \mathbf{W}\mathbf{x}^{(j)}_{in}||^2 \nonumber
    \end{align}
where $\beta > 0$ corresponds to the upper bound on the slope of the nonlinear activation function.

Using the label smoothness expression computed with input features to the layer, and gathering all the positive constants as $c= \beta\frac{c_{out}}{c_{in}}$, we obtain
\begin{align}
    \mathbf{y}^\top\mathcal{L}_{out}\mathbf{y} &\leq c\; ||\mathbf{W}||^2_2\; \mathbf{y}^\top\mathcal{L}_{in}\mathbf{y}.
\end{align}
\end{proof}

\subsection{Grad-CAM}

Gradient weighted Class Activation Maps (Grad-CAM) is a technique to visualize the reason behind the decision of a convolutional neural network based model. As convolutional layers naturally retain spatial
information which is lost in fully-connected layers, so, the last convolutional layers contains optimal high-level semantics and detailed spatial information. The neurons in these layers are responsible for semantic class-specific information in the image. The gradient information flowing into the last convolutional layer of the CNN is used in Grad-CAM to compute the importance of each neuron for a particular decision of interest.

Let the last layer produce $N$ feature maps, $A^n \in R ^{u \times v} $ with each element indexed by $i, j$. So $F^n_{i,j}$ refers to the activation at location $(i, j)$ of the feature map $F^k$ . The class discriminative localization map Grad-CAM $L_{Grad-CAM}^{c}$ for any class $c$ is computed performing a weighted summation of forward activation maps $F^{n}$  and following it by a ReLU as in (\ref{eqn:grad_cam}). To compute the weights $\alpha_{n}^{c}$, first, the gradient of the score for class $c, y^c$ (before the softmax), with respect to feature map activations $F^n$ of a convolutional layer, i.e. $\frac{\delta y^c}{\delta F^k}$ is obtained. The neuron weights $\alpha_{n}^{c}$ are computed  by global average pooling of these gradients. Finally, the Grad CAM is computed using the following equation (\ref{eqn:grad_cam}).

\begin{equation}
    \label{eqn:grad_cam}
    L_{Grad-CAM}^{c}=ReLU(\sum_n \alpha_{n}^{c} F^n )
\end{equation}

\subsection{Dynamic Time Warping}
\label{sec:DTW}
Dynamic time warping (DTW) \cite{masood2014dynamic} is a well-known technique to find an optimal
alignment between two given (time-dependent) sequences. DTW allows many-to-one comparisons to create the best possible alignment, exploiting temporal distortions, unlike Euclidean, which allows one-to-one point comparison. This is why DTW is efficient in computing similarity between two variable-length arrays or time sequences. Intuitively, the sequences are warped in a nonlinear fashion to match each other. In the action recognition task, we have action sequences of variable time length. Consequently, the features extracted also preserve the time localization as a property of STGCN. Thus, in this work, we use DTW to compute the similarity between two temporal features extracted using STGCN. DTW is computed using the following equation \ref{eqn:dtw}.  $dtw(i, j)$ is the minimum warp distance of two time series of lengths i and j. Any element $dtw(i,j)$  in the accumulated matrix indicates the dynamic time warping distance between series $A_{1:i}$ and $B_{1:j}$. 
 
  \begin{equation}
  \label{eqn:dtw}
  \begin{split}
        DTW(i,j) & =dist(a_i,b_j)+ \\
        & \min(DTW(i-1,j),DTW(i,j-1),DTW(i-1,j-1))
    \end{split}
 \end{equation}

\subsection{Spectral analysis of the skeleton graph}

\begin{figure}[!ht]
\begin{center}
    \frame{\includegraphics[width=.6\linewidth]{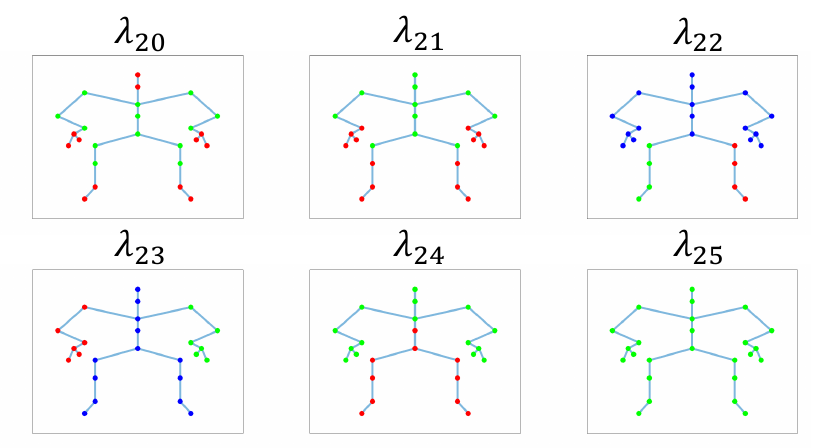}}
\end{center}
    \caption{Eigen basis corresponding to the higher eigenvalues of the normalized adjacency matrix. Here,  Green, red and blue dots  denote the positive values of the bases, the negative value and zero respectively  }
\label{fig:eigenbases_adjacency}
\end{figure}

\begin{figure}[htbp]
\centering
\centering
\includegraphics[width=.6\linewidth]{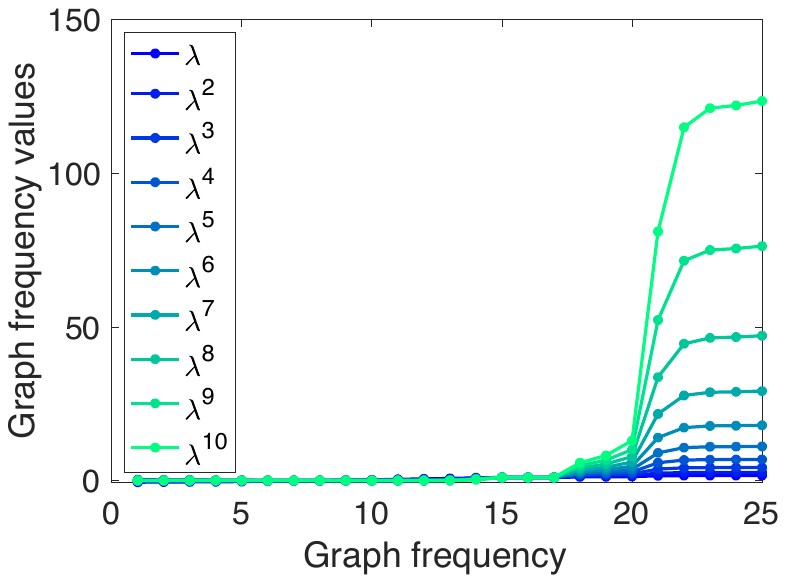}
   \caption{ The layers of STGCN correspond to learning graph filters that are based on polynomial powers of the eigenvalues. We show here the eigenvalues and their corresponding values for different exponents. We see that higher-degree polynomials are able to capture sufficiently the spectrum of human actions.
   }
\label{fig:energyspectrum_freq_polyfilter_powers}
\end{figure}
% \clearpage

\FloatBarrier
\subsection{Results}
\subsubsection{Experimental set-up}
We divided the actions in NTU-RGB60 in three super-classes corresponding to actions involving the upper body, lower body, and full body. Table \ref{tab:ntu60_actioncategory} shows the actions corresponding to actions involving the upper body, lower body, and full body.

\begin{table*}[h]
\centering
\begin{tabularx}{\textwidth}{|l|X|}
\textbf{Category}:               &  \textbf{Actions} \\
Upper body :          & 'drink water', 'eat meal or snack', 'brushing teeth',  'brushing hair', 'drop', 'pickup', 'throw', 'clapping', 'reading', 'writing', 'tear up paper', 'wear jacket', 'take off jacket',  'wear a shoe', 'take off a shoe', 'wear on glasses',  'take off glasses', 'put on a hat or cap', 'take off a hat or cap',  'cheer up', 'hand waving', 'reach into pocket', 'make a phone call\_answer phone', 'playing with phone\_tablet',  'typing on a keyboard', 'pointing to something with finger',  'taking a selfie', 'check time (from watch)', 'rub two hands together', 'nod head or bow', 'shake head',  'wipe face', 'salute', 'put the palms together',  'cross hands in front (say stop)', 'sneeze or cough', 'touch head (headache)', 'touch chest (stomachache or heart pain)', 'touch back (backache)', 'touch neck (neckache)',  'nausea or vomiting condition', 'use a fan (with hand or paper) or feeling warm', 'punching or slapping other person', 'pushing other person', 'pat on back of other person', 'point finger at the other person',  'hugging other person', 'giving something to other person', 'touch other persons pocket', 'handshaking' \\
Lower body:               & 'kicking something', 'hopping (one foot jumping)', 'jump up', 'kicking other person', 'walking towards each other', 'walking apart from each other' \\

Full body:               & 'sitting down', 'standing up (from sitting position)', 'staggering', 'falling' \\
\end{tabularx}
\caption{NTU-RGB60 Upper body, lower body and full body action category}
\label{tab:ntu60_actioncategory}
\end{table*}

\subsubsection{Results on other dataset and advanced network}
\begin{figure*}[!ht]
\begin{center}
\includegraphics[width=.5\linewidth]{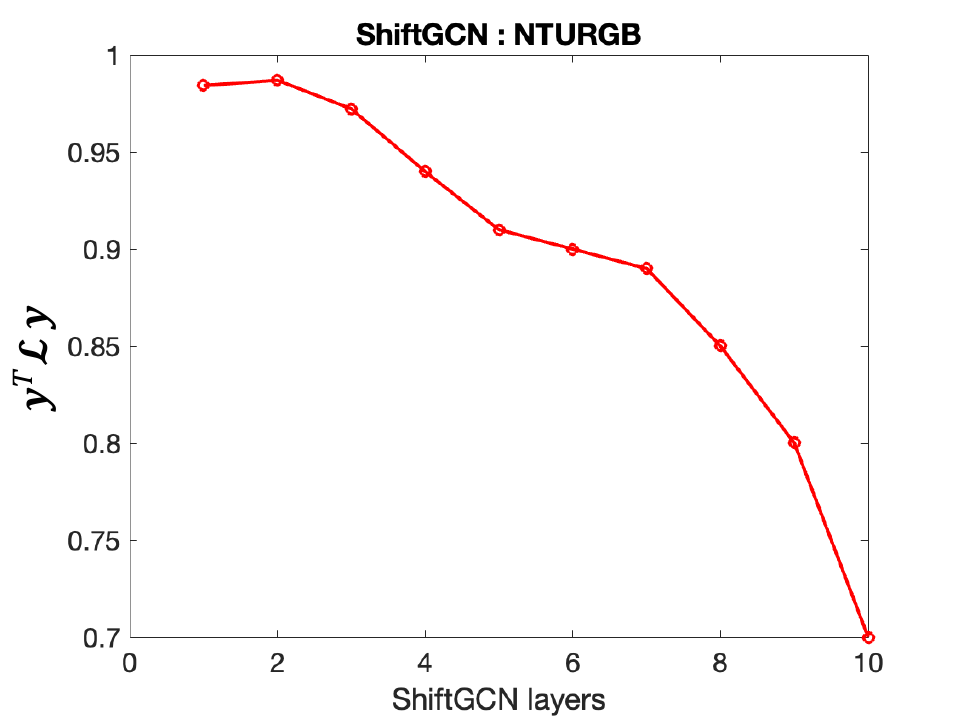} 
\end{center}
    \caption{Smoothness of labels on the manifold induced by the ShiftGCN \cite{cheng2020shiftgcn} layer mappings in a trained model. Intuitively, a lower value in $ \mathbf{y}^\top\mathcal{L}_{out}\mathbf{y}$ corresponds to the features belonging to a particular class having neighbors from the same class. This visualization of label smoothness for the ShiftGCN network trained on NTU-RGB 60 dataset.  We emphasize that, though the smoothness is displayed per class, the \textit{Dataset Graph}(NNK) is constructed using the features corresponding to all input action data points. We observe that the model follows a gradually decreasing pattern in $ \mathbf{y}^\top\mathcal{L}_{out}\mathbf{y}$ over the layers unlike STGCN. }
\label{fig:shiftgcn}
\end{figure*}

\paragraph{Results on ShiftGCN}
 Shift graph convolutional network (Shift-GCN) comprises shift graph operations and lightweight point-wise convolutions, where the shift graph operations provide flexible receptive fields for both spatial and temporal graphs. There are two modes of combining the shift operation with point-wise convolution, first, Shift-Conv and Shift-ConvShift. 
 To provide insights into the representations of the intermediate layers of the ShiftGCN network, we use our geometric analysis of the representation using the DTW-based NNK method described in sec~\ref{sec:knn_labelsmoothness}. 
Fig.~\ref{fig:shiftgcn}  shows the label smoothness over the layers of ShiftGCN demonstrating that our proposed DTW-based NNK method for geometric analysis of the representation can be used for any network dealing with spatiotemporal data. ShiftGCN shows a gradual fall in the Laplacian quadratic ($ \mathbf{y}^\top\mathcal{L}_{out}\mathbf{y}$) over the layers, unlike STGCN where we see a flat region in the initial layers and a sudden fall after layer $8$. The possible reason behind the gradual change in the label smoothness from the initial layers is the shift operation in the spatial and temporal domains. These shift operators allow ShiftGCN to learn global features starting from the initial layers showing a gradual increasing pattern in the label smoothness.

\begin{figure*}[!ht]
\begin{center}
\includegraphics[width=0.5\linewidth]{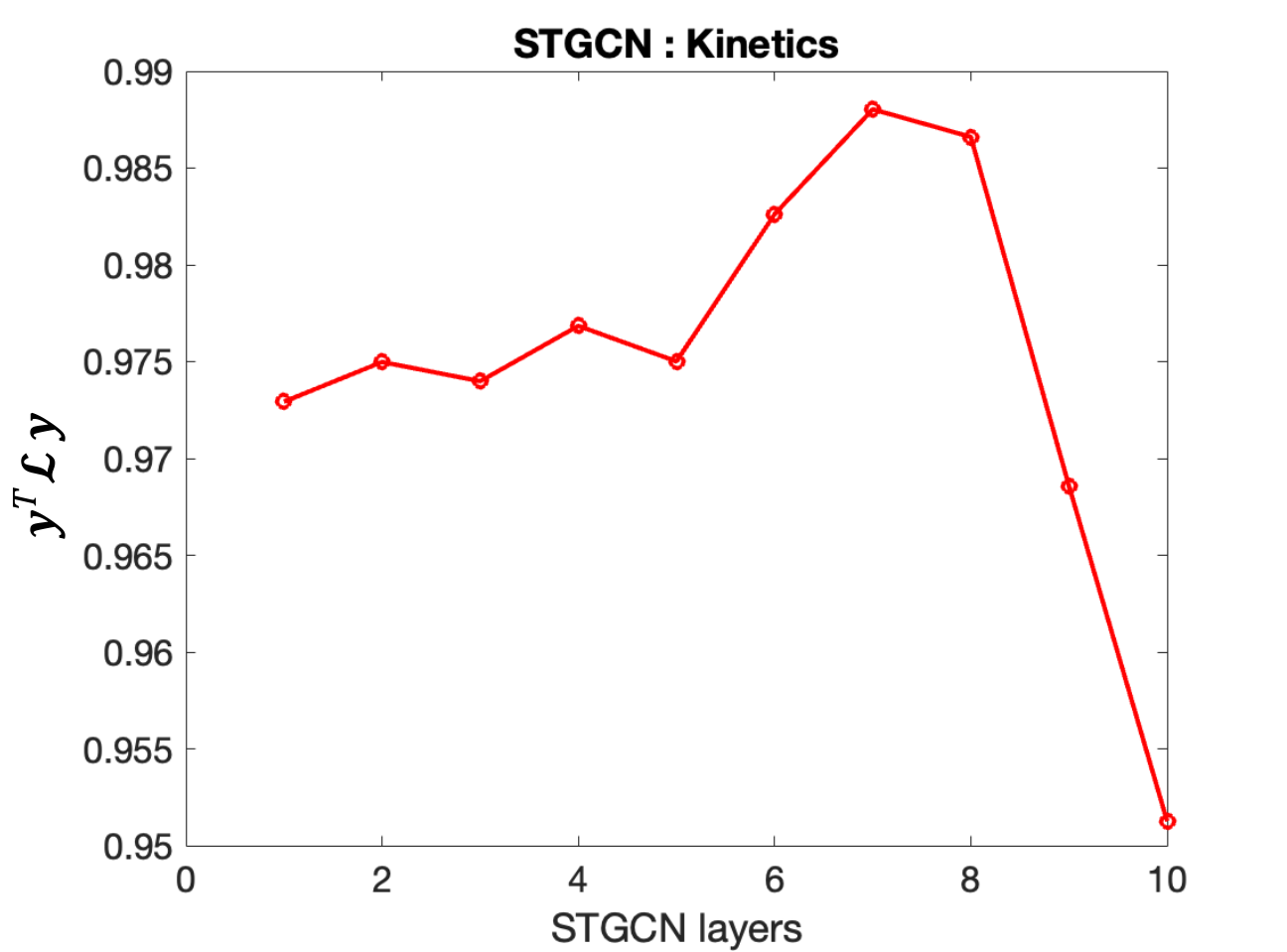} 
\end{center}
    \caption{This visualization of layer smoothness for the STGCN network trained on Kinetics 400 dataset \cite{kay2017kinetics}. Smoothness of labels on the manifold induced by the STGCN layer mappings in a trained model. Intuitively, higher smoothness corresponds to the features belonging to a particular class having neighbors from the same class.  We emphasize that, though the smoothness is displayed per class, the \textit{Dataset Graph}(NNK) is constructed using the features corresponding to all input action data points. We observe that the model follows a similar trend where the Laplacian quadratic $ \mathbf{y}^\top\mathcal{L}_{out}\mathbf{y}$ is flat in the initial layers (indicative of no class-specific learning) and decrease in value in the later layers (corresponding to discriminative learning}
\label{fig:kinetics}
\end{figure*}

\paragraph{Results on STGCN network trained on dataset Kinetics 400}
Deepmind Kinetics human action dataset \cite{kay2017kinetics} contains around 300, 000 video clips retrieved from YouTube with an average length of around 10 seconds. The videos cover as many as 400 human action classes, ranging from daily activities, sports scenes, to complex actions with interactions. Skeleton location
of 18 joints in 2D coordinates (X, Y ) format on each frame is extracted using
OpenPose \cite{osokin2018real}. The STGCN is trained using 240, 000 skeleton data samples, no background information was used. Fig.~\ref{fig:kinetics}  shows the label smoothness over the layers of STGCN trained on Kinetics dataset. We see similar pattern i.e. sudden drop in the Laplacian quadratic ($ \mathbf{y}^\top\mathcal{L}_{out}\mathbf{y}$) starting from layer $8$. This again implies that the last few layers learn particular information related the specific actions present in the dataset. We notice a relatively small change in the label smoothness over the layers of STGCN. As there are many actions in the Kinetics dataset which deal with different objects for example 'eating donuts', 'eating hotdog', 'eating burger', it is very difficult to differentiate the action without the object information. This could be one of the reason behind poor performance of STGCN on Kinetics dataset. However, our proposed NNK based embedding geometry understanding method can analyze STGCN network trained on any dataset. 

\FloatBarrier

\subsubsection{L-STG-GradCAM }

\begin{figure*}[!ht]
\begin{center}
\includegraphics[width=1\linewidth]{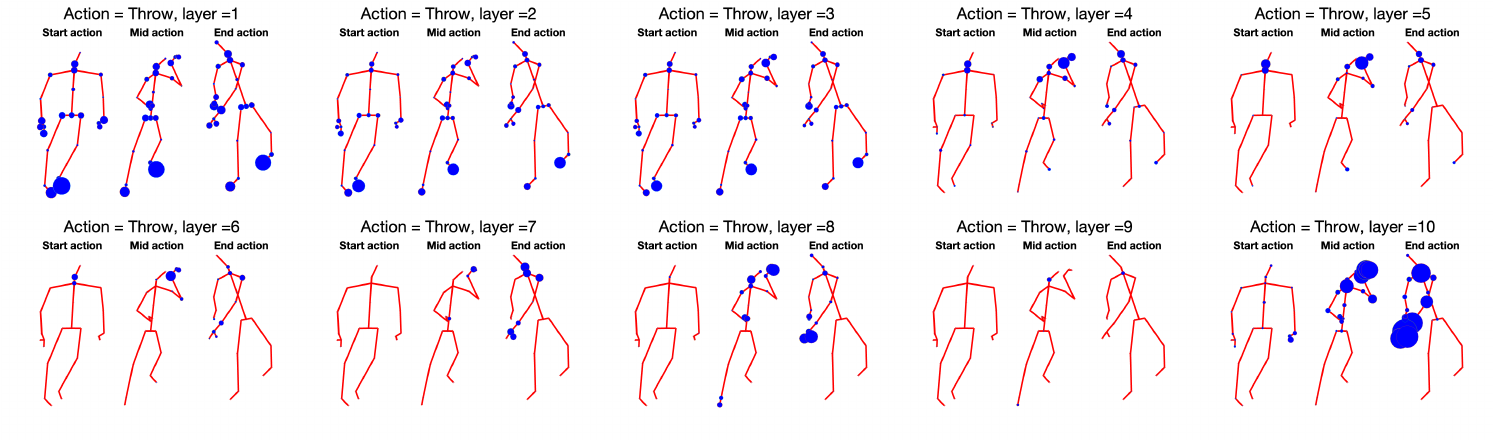} 
\end{center}
    \caption{L-STG-GradCAM  visualization of spatiotemporal node importance for action class \emph{Throw} of a trained STGCN network used in experiments. The size of the blue bubble denotes the relative importance of the node in a layer for prediction by the final softmax classifier and is scaled to have values in $[0, 1]$ at each layer. We observe that the localization of the action as observed using the L-STG-GradCAM is evident only in later layers while initial layers have no class-specific influence. 
The visualizations allow for transparency in an otherwise black-box model to explain any class prediction. Our approach is applicable to any STGCN model and is not affected by the model size, optimization strategy, or dataset used for training.
}
\label{fig:stggradcam_throw}
\end{figure*}

We consider a human action classification task using skeleton data to evaluate the performance of STG-Grad CAM extended for each layer. The $10$-layer STGCN model \cite{yan2018spatial} is used for evaluation. \autoref{fig:stggradcam_throw} shows the layerwise variation of joint importance for the action \textit{'Throw'} for three time-slices. The node's size in the S-graph denotes the degree of importance of the body joint at that time point for the final prediction. For the action \emph{Throw}, which mostly involves upper body parts, we notice in the figure that the initial layers (up to layer $8$) have very weak, if any, GradCAM localization corresponding to the action. In contrast, the last three STGCN layers show explicit node importance \emph{heatmap} where the one hand and the back joints are relatively more active, indicative of the action.

\begin{figure*}[h]
\begin{center}
\includegraphics[width=1\linewidth]{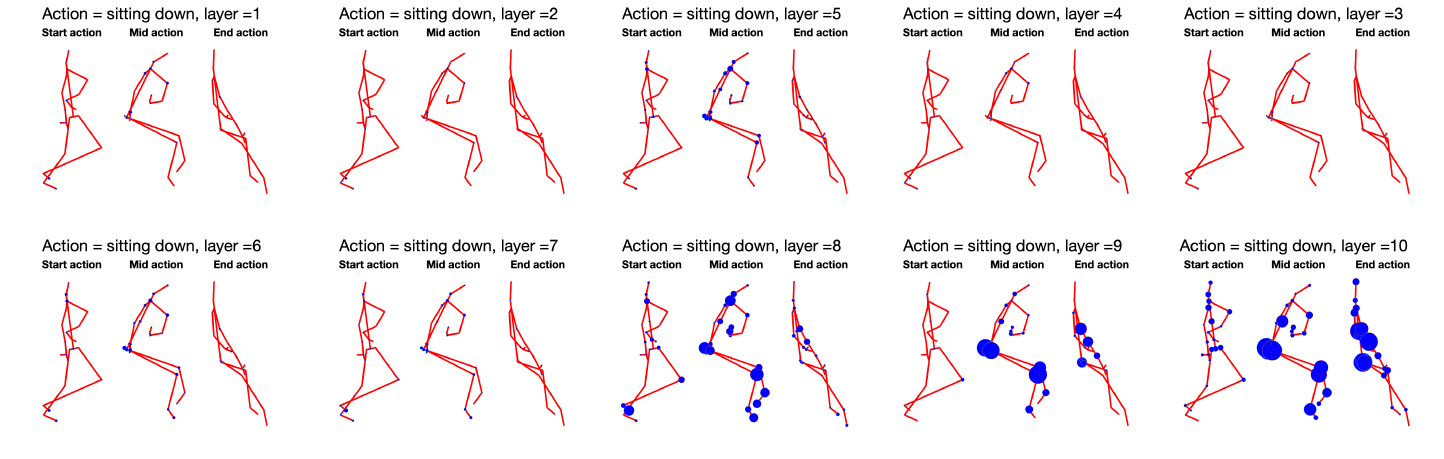}  
\end{center}
    \caption{L-STG-GradCAM  visualization of spatiotemporal node importance for action class \emph{Sitting down} of a trained STGCN network used in experiments. The size of the blue bubble denotes the relative importance of the node in a layer for prediction by the final softmax classifier and is scaled to have values in $[0, 1]$ at each layer. We observe that the localization of the action as observed using the L-STG-GradCAM is evident only in later layers while initial layers have no class-specific influence. 
% This observation of early layers capturing general human  motion is further reinforced in our geometric analysis using label smoothness.
The visualizations allow for transparency in an otherwise black-box model to explain any class prediction. Our approach is applicable to any STGCN model and is not affected by the model size, optimization strategy, or dataset used for training.
}
\label{fig:stggradcam_sitdown}
\end{figure*}

 \autoref{fig:stggradcam_sitdown} shows the layerwise variation of joint importance for the action \textit{'Sitting down'} for three time-slices. For the action \emph{Sitting down}, which mostly involves upper body parts, we notice in the figure that the initial layers (up to layer $8$) have very weak, if any, GradCAM localization corresponding to the action. In contrast, the last three STGCN layers show explicit node importance \emph{heatmap} where all the joints are relatively more active in the middle and end frame, indicative of the action spatially and temporally.

\FloatBarrier

% \subsection{Noisy samples}
% \begin{figure*}[h]
% \begin{center}
% \includegraphics[width=1\linewidth]{images/noisy_samples.pdf}  
% \end{center}
% \caption{Impact of noise added to the input sequence on the label smoothness observed using the corresponding features obtained with the model. We show the impact of noise on 12 actions that are well detected by STGCN. We observe that in actions where the label smoothness had a steady trend (non-increasing) remained mostly unaffected by the addition of noise to the input action sequence. However, actions, where the label smoothness increased before decreasing, were affected by noise. This implies that the features learned in the early layers for these actions are not robust and the addition of noise allows us to see the regularized manifold induced in these layers. }
% \label{fig:noisy12}
% \end{figure*}
\FloatBarrier

\subsection{Transfer Performance}
We divided the actions in NTU-RGB61-120 in three super-classes corresponding to actions involving the upper body, lower body, and full body. Table \ref{tab:ntu120_actioncategory} shows the actions corresponding to actions involving the upper body, lower body, and full body.

\begin{table*}[h]
\centering

\begin{tabularx}{\textwidth}{|l|X|}
\textbf{Category}:               &  \textbf{Actions} \\
Upper body :          & 'put on headphone',	'take off headphone',	'shoot at the basket',	'bounce ball',	'tennis bat swing',	'juggling table tennis balls',	'hush (quite)',	'flick hair',	'thumb up',	'thumb down',	'make ok sign',	'make victory sign',	'staple book',	'counting money',	'cutting nails',	'cutting paper (using scissors)',	'snapping fingers',	'open bottle',	'sniff (smell)',	'toss a coin',	'fold paper',	'ball up paper',	'play magic cube',	'apply cream on face',	'apply cream on hand back',	'put on bag',	'take off bag',	'put something into a bag',	'take something out of a bag',	'open a box',	'shake fist',	'throw up cap or hat',	'hands up (both hands)',	'cross arms',	'arm circles',	'arm swings',	'yawn',	'blow nose',	'hit other person with something',	'wield knife towards other person',	'grab other person’s stuff',	'shoot at other person with a gun',	'high-five',	'cheers and drink',	'take a photo of other person',	'whisper in other person’s ear',	'exchange things with other person',	'support somebody with hand',	'finger-guessing game (playing rock-paper-scissors)' \\
Lower body:               & butt kicks (kick backward)', 'cross toe touch', 'side kick', 'step on foot' \\

Full body:               & 'squat down', 'move heavy objects', 'running on the spot', 'stretch oneself', 'knock over other person (hit with body)', 'carry something with other person', 'follow other person'
 \\
\end{tabularx}
\caption{NTU-RGB120 Upper body, lower body and full body action category}
\label{tab:ntu120_actioncategory}
\end{table*}

% \begin{figure*}[!ht]
% \centering
% \includegraphics[width=1\linewidth]{images/nturgb120_label_Smoothness_actwise.pdf}    
% \caption{Smoothness of labels of the NTURGB120 dataset on the manifold induced by the STGCN layer mappings in a trained model on NTURGB60 . Intuitively, a lower value in smoothness corresponds to the features belonging to a particular class having neighbors that are from the same class. We divide the actions in NTU-RGB120 into three super-classes~(Upper body (\textbf{Left}), Full body (\textbf{Middle}), Lower body (\textbf{Right})) and present smoothness with respect to each action in the grouping. We emphasize that, though the smoothness is displayed per class, the $k$-NN graph is constructed using the features corresponding to all input action data points. We observe that the model follows a similar trend where the smoothness of labels is flat in the initial layers (indicative of no class-specific learning) and decrease in value in the later layers (corresponding to discriminative learning). There exist outliers to this trend~(e.g., in upper body group \emph{put on headphone}) where the smoothness increases in intermediate layers. This implies that the representations for these actions are affected by features from other actions to accommodate for learning other classes.}
% \label{fig:ntu120_lsmoothnness_actwise}
% \end{figure*}

% \FloatBarrier
\begin{figure}[h!]
\centering
\includegraphics[width=.5\linewidth]{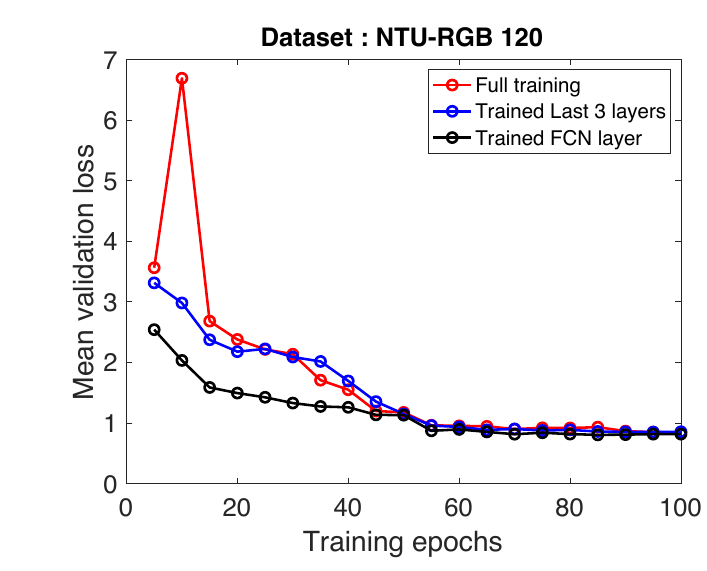}    
\caption{Validation loss (lower is better) on NTU-RGB120 test dataset using a $10$-layer STGCN network. We show here the validation loss achieved by the model when training from scratch along with that obtained with transfer learning by fine-tuning a model trained on NTU-RGB60 on fewer layers. We see that the model is able to transfer effectively as predicted by our label smoothness analysis.}
\label{fig:ntu120_valloss}
\end{figure}

 \FloatBarrier
% \clearpage

\section{Limitations}

% Our focus in this work is to introduce a data-driven framework for improving the understanding of STGCN networks. 
The analysis is based on features learned in a trained network and we foresee no issues while scaling up to larger models and datasets, but testing this remains open. We note that our approach can be applied to any spatiotemporal data and model, but the analysis in this work was limited to skeleton-based human action recognition. Although our work here makes some empirical observations, we note that our approach is amenable to theoretical study using spectral and graph signal concepts. We plan to investigate these ideas and ultimately develop an understanding that can lead to better design, training, and transfer of spatiotemporal models.
\FloatBarrier
\clearpage
\end{document}